\documentclass[a4paper,11pt]{article}
\pdfoutput=1 

\usepackage{jcappub} 

\usepackage{float}
\usepackage[T1]{fontenc} 
\usepackage{graphicx}

\newcommand{\kms}{\,km\,s$^{-1}$\,Mpc$^{-1}$}

\title{\boldmath Hubble Tension and Dark Energy in Teleparallel Gauss-Bonnet Gravity: New Constraints from DESI BAO, Pantheon$^+$ and Hubble Data}

\author[a,b]{Santosh V. Lohakare,}
\author[b,c,1]{S. K. Maurya,\note{Corresponding author}}
\author[b]{Aaisha  Al Qassabi,} 
\author[d]{B. Mishra}

\affiliation[a]{Department of Mathematics, School of Advanced Sciences, Vellore Institute of Technology, Vellore, 632014, Tamilnadu, India}
\affiliation[b]{Department of Mathematical and Physical Sciences, College of Arts and Sciences,\\ University of Nizwa, P.O. Box 33, Nizwa 616, Sultanate of Oman}
\affiliation[c]{Research Center of Astrophysics and Cosmology, Khazar University, Baku, AZ1096, 41 Mehseti Street, Azerbaijan}
\affiliation[d]{Department of Mathematics, Birla Institute of Technology and Science-Pilani, Hyderabad Campus, Jawahar Nagar, Kapra Mandal, Medchal District, Telangana 500078, India}

\emailAdd{lohakare.santosh@vit.ac.in}
\emailAdd{sunil@unizwa.edu.om}
\emailAdd{aaishakhalifa@unizwa.edu.om}
\emailAdd{bivu@hyderabad.bits-pilani.ac.in}

\abstract{We explore the cosmological dynamics of a teleparallel Gauss-Bonnet gravity model defined by the torsion scalar $T$ and the torsion-based Gauss-Bonnet invariant $T_{\mathcal{G}}$, deriving modified Friedmann equations for a flat FLRW Universe and corresponding linear scalar perturbation equations. Using a numerical approach, we solve these equations for pressureless matter, predicting the redshift evolution of the Hubble parameter $H(z)$. Using a Bayesian Markov chain Monte Carlo analysis based on late-time observations from Cosmic Chronometers, Pantheon$^+$ without SH0ES, and DESI BAO Data Release 1 and Data Release 2, we constrain the model parameters and show that this class of $f(T,T_\mathcal{G})$ cosmologies can mimic an effective dark-energy component without introducing an explicit cosmological constant. We further examine the scalar perturbation sector of the posterior-supported cosmological branch and find that the solutions considered in this work remain well-behaved within the adopted perturbative treatment. The model yields a present-day effective equation-of-state parameter in the range $\omega_{\rm eff}(z=0)\approx -0.664$ to $-0.693$, consistent with late-time observations, and shifts the inferred value of $H_0$ toward $69$--$71.5\ {\rm km\,s^{-1}\,Mpc^{-1}}$, suggesting a partial alleviation, though not a complete resolution, of the Hubble tension.}

\begin{document}
\maketitle
\flushbottom

\section{Introduction} \label{SEC I}
    Astronomical observations, including Type Ia Supernovae (SNe Ia) as standard candles \cite{Riess_1998_116, Perlmutter_1998_517, Kowalski_2008_686}, cosmic microwave background (CMB) temperature anisotropies \cite{Ade_2016_594a, Jarosik_2011_192, Komatsu_2009_180}, weak gravitational lensing \cite{Benjamin_2007_381, Fu_2008_479, Amendola_2008_04_013}, baryon acoustic oscillations (BAO), large-scale matter distribution \cite{Tegmark_2004_69, Cole_2005_362, Eisenstein_2005_633, Percival_2010_401, Blake_2012_425}, and high-redshift galaxy clustering \cite{Bunker_2010_409}, robustly confirm the accelerating expansion of the Universe. Within general relativity (GR), this acceleration is attributed to dark energy, often modeled as a cosmological constant ($\Lambda$) with a constant equation of state (EoS) parameter $\omega_{\Lambda} = -1$ \cite{Peebles_2003_75}, forming the basis of the $\Lambda$CDM model alongside cold dark matter (CDM). However, $\Lambda$CDM faces theoretical challenges like the fine-tuning and cosmic coincidence problems \cite{Weinberg_1989_61, Carroll_2001_4}, as well as observational tensions, such as the Hubble constant ($H_0$) discrepancy \cite{DiValentino_2021_38, DiValentino_2021_131, Montani_2026_113, Dainotti_2021_912, Dainotti_2025_48, Dainotti_2022_10, DeSimone_2025_45, Navone_2511.16130, Valletta_2512.19568, Montani_2024_44, Fazzari_2026_49} and the matter fluctuation amplitude ($S_8$) tension, though recent KiDS-Legacy results \cite{Wright_2025_2503.19441} suggest the latter is alleviated ($ S_8 = 0.815^{+0.016}_{-0.021} $, $ 0.73\sigma $ from Planck) \cite{Perivolaropoulos_2022_95, Valentino_Cosmoverse_white_paper}, prompting exploration of alternatives like dynamic dark energy within GR or modified gravity theories \cite{Tsujikawa_2010_800, Caldwell_1998_80, COPELAND_2006_15, Padmanabhan_2002_66}. Precise measurements of the dark energy EoS and its evolution remain critical to unraveling its nature \cite{COPELAND_2006_15, Frieman_2008_46, Weinberg_2013_530}.

    Recent cosmological findings from the Dark Energy Spectroscopic Instrument (DESI)~\cite{Adame_2024_DESI_collaboration} have marked a transformative advancement in our understanding of cosmic evolution. The DESI Stage IV survey significantly refines cosmological constraints through precise measurements of galaxy clustering, quasar distributions, and Lyman-$\alpha$ forest observations across extensive sky coverage and redshift ranges. The first year of DESI observations has yielded robust measurements of both transverse comoving distances and Hubble rates, derived from a sample exceeding six million extragalactic objects, the largest collected for such analyses to date. Initial results from the DESI BAO data release one (DR1) demonstrate remarkable consistency with the flat $\Lambda$CDM paradigm, reporting a matter density parameter $\Omega_{\mathrm{m}0} = 0.295 \pm 0.015$. When combined with Big Bang Nucleosynthesis priors and cosmic microwave background acoustic scale measurements, these data constrain the Hubble constant to $H_0 = 68.52 \pm 0.62$\kms. Further integration with Planck CMB anisotropy data and ACT/Planck CMB lensing measurements yields even tighter constraints: $\Omega_{\mathrm{m}0} = 0.307 \pm 0.005$ and $H_0 = 67.97 \pm 0.38$\kms~\cite{Adame_2024_DESI_collaboration}. These groundbreaking results not only validate the current cosmological standard model but also provide an exceptionally precise observational foundation for testing alternative theories of cosmic acceleration and large-scale structure formation.

    The phenomenon of late-time accelerated cosmic expansion has been extensively investigated through modifications to gravitational theory, offering an alternative to the dark energy paradigm. Within the framework of GR, such modifications are not feasible \cite{Lue_2006_423, Nojiri_2007_04}. To address this limitation, extensions to the geometrical structure of the Einstein-Hilbert action have been proposed to account for cosmic expansion. A prominent class of modified gravity theories emerges from extending the Einstein-Hilbert Lagrangian by incorporating additional curvature or matter-related terms. This generalized approach gives rise to several distinct formulations, including $f(R)$ (with $R$ denoting the Ricci scalar), $f(T)$ (with $T$ the torsion scalar), $f(Q)$ (with $Q$ the non-metricity scalar), $f(\mathcal{G})$ (with $\mathcal{G}$ the Gauss–Bonnet curvature term), and Lovelock gravity, among others (see \cite{Starobinsky_1980_91, De_Felice_2010_13, Fernandes_2022_39, Lovelock_1971_12, Cai_2016_79, Heisenberg_2023_1066_Review}). These theoretical frameworks have garnered significant interest within the scientific community due to their rich and diverse cosmological implications, including their potential to address key challenges in modern cosmology such as late-time cosmic acceleration, the Hubble tension and the nature of dark energy \cite{Vagnozzi_2022_36, Vagnozzi_2020_102, Ilyas_2020_09, Lohakare_2024_MNRAS, Lohakare_2025_ApJ, Bamba_2012_342}. In the teleparallel formulation of gravity, higher curvature corrections, such as the Gauss-Bonnet invariant $T_\mathcal{G}$, can be incorporated, leading to actions that include higher torsion modifications \cite{Boulware_1985_55, Wheeler_1986_268, Antoniadis_1994_415, Nojiri_2005_71}. Notably, the torsion-based Gauss-Bonnet invariant $T_\mathcal{G}$ has been derived without relying on the Weitzenb\"{o}ck connection, demonstrating equivalence to the Gauss-Bonnet term $\mathcal{G}$ \cite{Kofinas_2014_90_084044}. This development has given rise to a compelling class of modified gravitational theories, termed $f(T, T_\mathcal{G})$ gravity \cite{Kofinas_2014_90_084044, Kofinas_2014_90_084045}.

    Another avenue of modified gravity involves coupling the torsion scalar $T$ with the trace of the energy-momentum tensor $\mathcal{T}$. This approach enables a unified description of cosmological phases, encompassing inflation, matter-dominated expansion, and late-time acceleration \cite{Harko_2014_2014_021, Duchaniya_2025_85}. Furthermore, extending $f(T)$ gravity by incorporating nonminimal torsion matter coupling into the action has proven effective in addressing the dark energy sector of the Universe \cite{Harko_2014_89}. This study focuses on a gravitational action combining the torsion scalar and the Gauss-Bonnet invariant, formulated within $f(T, T_\mathcal{G})$ gravity theories. These theories have been thoroughly explored across various contexts \cite{Kofinas_2014_90_084045, Chattopadhyay_2014_353, Capozziello_2016_76_629}, yielding significant insights on multiple scales. The present research investigates the cosmological dynamics of a specific subclass of $f(T, T_\mathcal{G})$ models, selected based on symmetry considerations. The primary objective is to evaluate the viability of this scenario as an alternative to the standard cosmological paradigm using late-time cosmic observations. In this work, a numerical methodology has been developed to predict the redshift evolution of the Hubble expansion rate. The $f(T, T_\mathcal{G})$ model emerges as a promising candidate for explaining the current cosmological epochs, effectively capturing the evolution of energy components over cosmic time and reinforcing its potential as a viable explanation for the observed accelerated expansion of the Universe. To assess the model comprehensively, the background cosmological dynamics are analyzed, and its feasibility is evaluated through Bayesian inference, employing Markov chain Monte Carlo (MCMC) analysis. This analysis incorporates late-time observational datasets, including Supernovae Ia (Pantheon$^+$), observational Hubble data (CC sample), DESI BAO Data Release 1 (DR1) and Data Release 2 (DR2) datasets. Additionally, a scalar perturbation analysis framework is introduced to examine the stability of the model. The results demonstrate that the model robustly represents the evolving energy components across cosmic time, lending credence to its role as a compelling alternative explanation for the observed cosmic acceleration.

    The novelty of the present work is not the introduction of a new $f(T, T_\mathcal{G})$ framework, but a phenomenological reassessment of a minimal subclass, $f(T, T_\mathcal{G})$, using updated late-time data including Pantheon$^+$, CC, and DESI DR1/DR2 BAO, together with an examination of the scalar perturbation sector. In that sense, the paper aims to refine the phenomenological status of this subclass rather than to claim a new gravitational formalism. It is useful to stress the scope of that claim more explicitly. The present work does not propose a new teleparallel Gauss--Bonnet framework; rather, it revisits a specific minimal subclass of $f(T,T_\mathcal{G})$ gravity in light of updated low-redshift observations and examines, in the same analysis, both the background expansion history and the scalar perturbation behavior of the cosmological branch supported by the data. In that sense, the contribution of this paper is mainly phenomenological. This distinction may seem modest, but it matters in practice: once Pantheon$^+$, Cosmic Chronometers, and the DESI DR1/DR2 BAO measurements are included, the allowed parameter region and the inferred late-time behavior can differ nontrivially from earlier studies based on older BAO compilations or more schematic data combinations. We also note that several earlier studies of $f(T,T_\mathcal{G})$ gravity focused either on formal aspects of the theory, dynamical-system behavior, or background evolution for different functional forms. By contrast, the aim here is narrower but more data-driven: to test whether the particular $f(T, T_\mathcal{G})$ model remains observationally viable in the DESI era and whether the corresponding late-time solutions remain acceptable in the scalar perturbation sector considered in this work.
    
    This study investigates late-time cosmic acceleration in a minimal $f(T, T_\mathcal{G})$ model by numerically solving the modified Friedmann equations and confronting the resulting expansion history with observational data. The article is structured into five sections: Section \ref{SEC II} establishes the geometrical and mathematical foundation of teleparallelism, formulates $f(T, T_\mathcal{G})$ gravity, and derives the general metric and field equations. Section \ref{SEC III} applies this framework to a cosmological context, developing $f(T, T_\mathcal{G})$ cosmology supported by observational datasets. Section \ref{SEC IV} builds on this model, conducting a scalar perturbation analysis to assess its stability. The article concludes in Section \ref{SEC conclusion} with a summary of results and a discussion of the implications for cosmological modeling and potential future directions.

\section{Mathematical Formalism of \texorpdfstring{$f(T, T_\mathcal{G})$}{} Gravity} \label{SEC II}
    In the teleparallel formulation of gravity, the fundamental dynamical variables are the tetrad field $e_a(x^\mu)$ and the connection one-forms $\omega^a_{\;\; b}(x^\mu)$, which together define parallel transport. Here, Greek indices $\mu, \nu, \ldots$ denote spacetime coordinates, while Latin indices $a, b, \ldots$ refer to the tangent (Lorentz) space. These fields can be expressed in coordinate components as $e_a = e^{\;\; \mu}_a \partial_\mu$ and $\omega^a_{\;\; b} = \omega^a_{\;\; b\mu} dx^\mu = \omega^a_{\;\; bc} e^c$, where the dual tetrad is defined as $e^a = e^a_{\;\; \mu} dx^\mu$.

    The commutation relations among the tetrad fields are given by
        \begin{equation}
            [e_{a},e_{b}]=C^{c}_{\,\,\,ab}e_{c}\,,
            \label{ghw}
        \end{equation}
    where the structure functions $C^c_{\;\; ab}$ are defined as
        \begin{equation}
            C^{c}_{\,\,\,ab}=e_{a}^{\,\,\,\mu} e_{b}^{\,\,\,\nu}(e^{c}_{\,\,\,\mu,\nu}-e^{c}_{\,\,\,\nu,\mu})
            \label{structurefun}\,,
        \end{equation}
    with a comma indicating partial differentiation with respect to the coordinate.

    We now define the torsion tensor in terms of tangent space components as
        \begin{equation}
            T^{a}_{\;\; bc} = \omega^{a}_{\;\; cb} - \omega^{a}_{\;\; bc} - C^{a}_{\;\; bc}\,,
        \end{equation}
    and the curvature tensor is given by
        \begin{equation}
            R^{a}_{\;\; bcd} = \partial_c \omega^{a}_{\;\; bd} - \partial_d \omega^{a}_{\;\; bc} + \omega^{e}_{\;\; bd} \omega^{a}_{\;\; ec} - \omega^{e}_{\;\; bc} \omega^{a}_{\;\; ed} - C^{e}_{\;\; cd} \omega^{a}_{\;\; be}.
            \label{curvaturebastard}
        \end{equation}

    To establish orthonormality, we employ the spacetime metric $g$ such that $g(e_a, e_b) = \eta_{ab}$, where $\eta_{ab} = \mathrm{diag}(-1,1,\ldots,1)$ denotes the Minkowski metric. This leads to the relation
        \begin{equation}
            \label{metrdef}
            g_{\mu\nu} = \eta_{ab} \, e^a_{\;\; \mu} \, e^b_{\;\; \nu},
        \end{equation}
    with Latin indices $a, b, \ldots$ raised and lowered using $\eta_{ab}$. 

    Finally, we introduce the contortion tensor, which captures the deviation from the Levi-Civita connection, defined as
        \begin{equation}
            \mathcal{K}_{abc} = \frac{1}{2} \left( T_{cab} - T_{bca} - T_{abc} \right) = -\mathcal{K}_{bac}.
        \end{equation}

    We now impose the condition of teleparallelism, namely the vanishing of the curvature tensor, $R^{a}_{\;\; bcd} = 0$, which is required to hold in all frames. One way to satisfy this condition is by adopting the Weitzenb{\"o}ck connection ($\tilde{\omega}^{\lambda}_{\;\; \mu\nu}$), which is defined purely in terms of the tetrad as
        \begin{eqnarray}
            \tilde{\omega}^{\lambda}_{\;\; \mu\nu} = e^{\;\; \lambda}_a \, \partial_\nu e^a_{\;\; \mu}.   
        \end{eqnarray}

    In the preferred frame (tangent-space components), this choice yields vanishing spin connection components: $\tilde{\omega}^{a}_{\;\; bc} = 0$.

    The Ricci scalar $R$, constructed from the standard Levi-Civita connection, can be related to the torsion scalar $T$ through the identity \cite{Aldrovandi_2013_TG_INTRO, Maluf_1994_35,  Arcos_2004_13}
        \begin{equation}
            e R = -e T + 2\, \partial_\mu \left( e\, T^{\nu\;\;\mu}_{\;\; \nu} \right),
            \label{ricciscalar}
        \end{equation}
    where $e = \det(e^a_{\;\; \mu}) = \sqrt{|g|}$, and the torsion scalar $T$ is defined by
        \begin{equation}
            T = \frac{1}{4} T^{\mu\nu\lambda} T_{\mu\nu\lambda} + \frac{1}{2} T^{\mu\nu\lambda} T_{\lambda\nu\mu} - T^{\nu}_{\;\; \nu\mu} T^{\lambda\;\;\mu}_{\;\; \lambda}.
            \label{Tscalar}
        \end{equation}

    In $f(T, T_\mathcal{G})$ gravity \cite{Kofinas_2014_90_084044, Kofinas_2014_90_084045}, the total modified gravitational action is given by
        \begin{equation}\label{eq.1}
            \mathcal{S}=\frac{1}{2\kappa^2} \int d^{4}x\,\, e\,\, f(T, T_\mathcal{G})\, .
        \end{equation}
        
    Having established the geometrical preliminaries, we now turn to the theoretical framework adopted in this work. In this study, we explore a gravitational framework grounded in the torsion scalar $T$ and the teleparallel Gauss-Bonnet invariant term $T_\mathcal{G}$. This approach generalizes the teleparallel equivalent of general relativity (TEGR) by incorporating higher-order corrections through $T_\mathcal{G}$, offering a compelling alternative to curvature-based modifications of gravity. We adopt the coupling constant $\kappa^2=8 \pi G$, where $G$ represents the Newtonian gravitational constant. In curvature-based gravity, the Gauss-Bonnet invariant is expressed as
        \begin{equation}
            \mathcal{G} \equiv R^2-4R^{a b} R_{a b}+R^{a b c d}R_{a b c d}    
        \end{equation}
    where $R$, $R^{a b}$, $R^{a b c d}$ denote the Ricci scalar, Ricci tensor, and Riemann tensor, respectively. In contrast, within the framework of torsion-based $f(T, T_{\mathcal{G}})$ gravity, the invariant $T_{\mathcal{G}}$ is defined as
        \begin{eqnarray}\label{eq.2}          T_\mathcal{G} =\Big(\mathcal{K}^{a_1}_{~~e a} \mathcal{K}^{e a_2}_{~~b} \mathcal{K}^{a_3}_{~~f c} \mathcal{K}^{f a_4}_{~~d}-2 \mathcal{K}^{a_1 a_2	}_{~~~~a} \mathcal{K}^{a_3}_{~~e b} \mathcal{K}^{e}_{~~f c} \mathcal{K}^{f a_4}_{~~d} \nonumber \\
            +2\mathcal{K}^{a_1 a_2	}_{~~~~a}  \mathcal{K}^{a_3}_{~~e b} \mathcal{K}^{e a_4}_{~~f} \mathcal{K}^{f}_{~c d}   +2\mathcal{K}^{a_1 a_2	}_{~~~~a} \mathcal{K}^{a_3}_{~~e b} \mathcal{K}^{e a_4}_{~~c,d} \Big)\delta^{a b c d}_{a_1 a_2 a_3 a_4}.
        \end{eqnarray}
    where $\delta^{a b c d}_{a_1 a_2 a_3 a_4}$ is the generalized Kronecker delta. This formulation encapsulates the geometric structure of torsion-based gravity, facilitating the exploration of cosmological dynamics distinct from curvature-based approaches.

    The gravitational field equations corresponding to the modified action \eqref{eq.1} are derived with respect to the tetrad formulation                      
        \begin{eqnarray}\label{eq.3}
            2(H^{[ac]b}+H^{[ac]b})_{,c} +2(H^{[ac]b}+H^{[ba]c}-H^{[cb]a}) \mathcal{C}^{d}_{\,\,\,\,dc} +(2H^{[ac]d}+H^{dca}) \mathcal{C}^{b}_{\,\,\,\,cd} \nonumber
            \\ +4 H^{[db]c} \mathcal{C}_{(dc)}^{\,\,\,\,\,\,\,\,\,a} +T^{a}_{\,\,\,\,cd} H^{cdb}-h^{ab} +(f-T f_T-T_\mathcal{G} f_{T_\mathcal{G}}) \eta^{ab}=0.
        \end{eqnarray}
    with
    \begin{eqnarray}
H^{abc}&=&f_T (\eta^{ac} \mathcal{K}^{bd}_{\,\,\,\,\,\,d}-\mathcal{K}^{bca}+f_{T_\mathcal{G}}\Big[\epsilon^{cprt}(2\epsilon^{a}_{\,\,\,dkf} \mathcal{K}^{bk}_{\,\,\,\,\,\,p}\mathcal{K}^{d}_{\,\,\,\,qr}+\epsilon_{qdkf} \mathcal{K}^{ak}_{\,\,\,\,\,\,p} \mathcal{K}^{bd}_{\,\,\,\,\,\,r}+\epsilon^{ab}_{\,\,\,\,\,\,kf} \mathcal{K}^{k}_{\,\,\,\,dp} \mathcal{K}^{d}_{\,\,\,\, qr}) \mathcal{K}^{qf}_{\,\,\,\,\,\,t} \nonumber \\ && + \epsilon^{cprt} \epsilon^{ab}_{\,\,\,\,kd} \mathcal{K}^{fd}_{\,\,\,\,\,\,p} (\mathcal{K}^{k}_{\,\,\,\,fr,t}-\frac{1}{2} \mathcal{K}^{k}_{\,\,\,\,fq} \mathcal{C}^{q}_{\,\,\,tr})+\epsilon^{cprt}\epsilon^{ak}_{\,\,\,\,\,\,df} \mathcal{K}^{df}_{\,\,\,\,\,\,p} (\mathcal{K}^{b}_{\,\,\, kr,t}-\frac{1}{2} \mathcal{K}^{b}_{\,\,\,\, kq} \mathcal{C}^{q}_{\,\,\,\,tr})\Big] \nonumber \\ &&+f_{T_\mathcal{G}} \mathcal{C}^{q}_{\,\,\,\,pt} \mathcal{K}^{bk}_{\,\,\,\,\,\,[q} \mathcal{K}^{df}_{\,\,\,\,\,\,r]}+\epsilon^{cprt} \epsilon^{a}_{\,\,\,\,kdf}\big[(f_{T_\mathcal{G}} \mathcal{K}^{bk}_{\,\,\,\,\,\,p} \mathcal{K}^{df}_{\,\,\,\,\,\,\,r})_{,t}\big]
\end{eqnarray}
    and
        \begin{equation*}
            h^{ab}=f_T \epsilon^{a}_{\,\,\,\,kcd} \epsilon^{bpqd} \mathcal{K}^{k}_{\,\,\,\,fp} \mathcal{K}^{fc}_{\,\,\,\,\,\,q}
        \end{equation*}
    where $f_T$ and $f_{T_\mathcal{G}}$ represent the partial derivatives of $f$ with respect to the torsion scalar $T$ and the Gauss-Bonnet invariant $T_\mathcal{G}$, respectively.

    To derive the field equations of $f(T, T_{\mathcal{G}})$, we consider an isotropic and homogeneous Friedmann--Lema\^{i}tre--Robertson--Walker (FLRW) spacetime, given by 
        \begin{equation} \label{eq.4}
            ds^{2}=-dt^{2}+a^{2}(t)(dx^{2}+dy^{2}+dz^{2}),
        \end{equation}
    where the scale factor $a(t)$ characterizes the uniform expansion of the Universe along spatial directions in a flat FLRW geometry. The diagonal tetrad for this spacetime is given by
        \begin{equation}\label{eq.5}
            e^{a}{}_{\mu}=\mathrm{diag}(1,\, a(t),\,\, a(t),\,\, a(t)) \, ,
        \end{equation} 
    with a determinant $e=\mathrm{det}(e^{a}{}_{\mu})=\sqrt{-g}=a^3(t)$. The dual tetrad is expressed as 
        \begin{equation} \label{eq.6}
            e_{\mu}{}^{a}=\mathrm{diag}(1,\, a^{-1}(t),\,\, a^{-1}(t),\,\, a^{-1}(t)) \, .
        \end{equation}
    
    In this context, the torsion scalar $T$ and the teleparallel Gauss-Bonnet invariant $T_\mathcal{G}$ are formulated in terms of the Hubble parameter $H=\frac{\dot{a}(t)}{a(t)}$, where a dot denotes differentiation with respect to cosmic time $t$. Specifically, they are given by
    \begin{eqnarray}
        T=6H^{2}\, , \hspace{2cm}  T_{\mathcal{G}}=24H^{2}(\dot{H}+H^{2})
    \end{eqnarray}

    At this stage, we emphasize that the present analysis is restricted to the spatially flat, homogeneous, and isotropic FLRW cosmological branch, for which the diagonal tetrad $e^a{}_\mu={\rm diag}(1,a,a,a)$ provides the standard background realization used throughout this work. Our aim is the late-time background and scalar-sector phenomenology of this branch, not a general analysis of all tetrad realizations in $f(T,T_\mathcal{G})$ gravity. We also note that local Lorentz/frame issues are known subtleties in extended teleparallel theories; in the present paper these issues do not invalidate the restricted FLRW treatment employed here, but they do delimit the scope of the claims being made. We further incorporate a matter action $\mathcal{S}_\mathrm{m}$, corresponding to an energy-momentum tensor $\mathcal{T}^{ab}$, modeled as a perfect fluid with energy density $\rho$ and pressure $p$. By varying the total action $\mathcal{S} + \mathcal{S}_\mathrm{m}$, we derive the field equations in the FLRW geometry, as established by Kofinas et al. \cite{Kofinas_2014_90_084045}
        \begin{eqnarray} 
            && f-12 H^2 f_{T} - T_{\mathcal{G}} f_{T_{\mathcal{G}}}+24 H^3 \dot{f}_{T_{\mathcal{G}}}=2 \kappa^2 \rho_\mathrm{m},  \label{Eq: first_field} \\
            && f-4(\dot{H}+3 H^2)f_{T}-4 H \dot{f}_{T}-T_{\mathcal{G}} f_{T_{\mathcal{G}}}+\frac{2}{3H} T_{\mathcal{G}} \dot{f}_{T_{\mathcal{G}}} +8 H^2 \ddot{f}_{T_{\mathcal{G}}}=-2 \kappa^2 p_\mathrm{m}.   \label{Eq: second_field}
        \end{eqnarray} 
    with
        \begin{eqnarray} \label{eq.9}
            f_T \equiv \frac{\partial f(T, T_\mathcal{G})}{\partial T},\,\,\,\,\,\,\,\,\,\,\,\,\, f_{T_\mathcal{G}} \equiv \frac{\partial f(T, T_\mathcal{G})}{\partial T_\mathcal{G}} \, .
        \end{eqnarray}
    
    Hereafter, we adopt the shorthand notation $f\equiv f(T, T_{\mathcal{G}})$ to streamline subsequent expressions. The standard matter density $\rho_\mathrm{m}$ is conserved independently, satisfying the continuity equation $\dot{\rho}_\mathrm{m} + 3H(\rho_\mathrm{m} + p_\mathrm{m}) = 0$. To construct a physically viable cosmological model, we compute the pressure $p$ and energy density $\rho$ for a suitable form of $f(T, T_\mathcal{G})$. The time derivatives of the partial derivatives are expressed as
        \begin{eqnarray}                        \dot{f}_T&=&f_{TT}\dot{T}+f_{TT_{\mathcal{G}}}\dot{T}_{\mathcal{G}} \, , \nonumber\\ \dot{f}_{T_{\mathcal{G}}}&=&f_{T T_\mathcal{G}}\dot{T}+f_{T_\mathcal{G} T_\mathcal{G}}\dot{T}_\mathcal{G} \, , \nonumber\\ \ddot{f}_{T_\mathcal{G}}&=&f_{TTT_{\mathcal{G}}}\dot{T}^2+2f_{{T T_\mathcal{G}} T_\mathcal{G}}\dot{T}\dot{T}_\mathcal{G}+f_{T_\mathcal{G} T_\mathcal{G} T_\mathcal{G}}\dot{T}_{\mathcal{G}}^2 +f_{T T_{\mathcal{G}}} \ddot{T} +f_{T_\mathcal{G} T_\mathcal{G}}\ddot{T}_{\mathcal{G}} ,\nonumber 
        \end{eqnarray}
    where $f_{TT}$, $f_{TT_\mathcal{G}}$, and similar terms represent higher-order partial derivatives of $f(T, T_\mathcal{G})$ with respect to $T$ and $T_\mathcal{G}$. These expressions facilitate the analysis of cosmological evolution within the $f(T, T_\mathcal{G})$ gravity framework.

\subsection{Numerical Results}
    The preceding analysis necessitates specifying the functional form of $f(T, T_\mathcal{G})$. In conventional $f(T)$ gravity, corrections typically involve powers of the torsion scalar $T$. However, in the context of $f(T, T_\mathcal{G})$ gravity, the teleparallel equivalent of the Gauss-Bonnet term $T_\mathcal{G}$ is of the same order as $T^2$, since it includes quartic torsion contributions. Given that both $T$ and $\sqrt{T^2 + \lambda_2 T_\mathcal{G}}$ scale similarly, a consistent modified gravity framework must incorporate both. Consequently, a minimal yet non-trivial model that departs from GR without introducing a new mass scale is defined by
        \begin{eqnarray}
            f(T, T_\mathcal{G}) = -T + \alpha \sqrt{T^2 + \beta T_\mathcal{G}}\, ,        
        \end{eqnarray}
    where $\alpha$ and $\beta$ are dimensionless coupling parameters. This model, despite its simplicity, is capable of yielding rich and distinctive cosmological behavior, particularly relevant at late times, highlighting the potential and flexibility of $f(T, T_\mathcal{G})$ gravity \cite{Kofinas_2014_31_175011, Lohakare_2023_39}. Notably, when $\beta = 0$, the theory reduces to the TEGR, which is equivalent to GR with a rescaled Newton's constant. The dynamical behavior of this case has been thoroughly investigated in the literature~\cite{Copeland_1998_57, Ferreira_1997_79, Chen_2009_2009_04} for the condition $\beta = 0$. Therefore, in the following, we restrict our attention to scenarios where $\beta \neq 0$. This choice is also convenient from the standpoint of cosmological scaling. In an FLRW background one has $T\propto H^2$ and $T_\mathcal{G}\propto H^2(\dot H+H^2)$, so that during standard power-law eras the $T_\mathcal{G}$ contribution is of the same effective order as $T^2$. For this reason, the model does not introduce an unrelated new scale at the level of the background equations. While the present work does not attempt a full early-Universe analysis including BBN and CMB anisotropies, this scaling behavior suggests that the model can accommodate the standard radiation and matter-dominated expansion regimes at the background level within an admissible region of parameter space.

    To derive the theoretical Hubble rate $H(z)$ for our model, we numerically solve the modified Friedmann equation~\eqref{Eq: first_field}. For the cosmological regime dominated by pressureless matter ($p_\mathrm{m} = 0$), the energy density evolves as
        \begin{equation}
            \rho_{\mathrm{m}}(z) = 3H_0^2 \Omega_{\mathrm{m}0} (1+z)^3,
        \end{equation}
    where $z$ is the cosmological redshift (defined via $a_0/a = 1+z$, with $a_0$ and $a$ representing the present-day and emission-time scale factors, respectively), and $\Omega_{\mathrm{m}0}$ denotes the current matter density parameter. Within this framework, the first Friedmann equation for our specific model takes the following form
       \begin{eqnarray} \label{Eq: main ode}
            & & \hspace{-0.35cm} - 3 {H_0}^2 \Omega_{\mathrm{m}0} (z+1)^3 + \frac{H^5}{\left((2 \beta +3) H^4 -2 \beta  (z+1) H^3 H'\right)^{3/2}}  \Bigg(-\sqrt{3} \alpha  \beta ^2 H (z+1)^2 {H'}^2 \nonumber\\ & & +H \Big((2 \beta +3) \Big(\sqrt{3} \alpha  (\beta -3) H^2 -\sqrt{3} \alpha  \beta ^2 H (z+1)^2 H'' +3 \sqrt{H^3 \left((2 \beta +3) H-2 \beta  (z+1) H'\right)}\Big) \Big) \nonumber\\ & & - \beta  (z+1) H' \Big(\sqrt{3} \alpha  (\beta -9) H^2 +6 \sqrt{H^3 \left((2 \beta +3) H-2 \beta  (z+1) H'\right)}\Big)\Bigg)=0 \, ,
        \end{eqnarray} 
    where the prime symbol ($^\prime$) denotes differentiation with respect to the redshift parameter $z$. The second-order nonlinear differential equation \eqref{Eq: main ode} for $H(z)$ requires two boundary conditions: (i) the present-day value $H(0) = H_0$, and (ii) the derivative constraint $H'(0) = \gamma$, where $\gamma$ is optimized via MCMC analysis. This data-driven approach enables rigorous exploration of the parameter space while satisfying observational constraints. The $\Lambda$CDM model is characterized by the Hubble function, defined as
        \begin{eqnarray}
            H_{\Lambda \mathrm{CDM}} = H_0 \sqrt{\Omega_{\mathrm{m}0} (1+z)^3 + \Omega_{\mathrm{DE}_0}}  \, ,
        \end{eqnarray}
    where $\Omega_{\text{DE}_0}$ representing the current value of the DE density parameter. For reproducibility, we note that the cosmological analysis is based on a custom numerical solver for Eq.~\eqref{Eq: main ode}, coupled to an MCMC likelihood pipeline. The model equation, boundary conditions, parameter priors, dataset combinations, and posterior-analysis tools used in this work are all specified explicitly in the text. The implementation can be made available upon reasonable request, and a public release may be prepared in the final publication stage if required.

\section{Cosmological Observations with Numerical Approach} \label{SEC III}
    We employ the \texttt{GetDist} library \cite{Lewis_getdist}, a widely used toolkit for producing high-quality one and two-dimensional posterior distribution plots. The parameter space is explored through MCMC sampling with the \texttt{emcee} Python package \cite{Foreman_Mackey_2013_emcee}. The resulting chains are then analyzed with the GetDist Python interface \cite{Lewis_getdist} to extract parameter constraints and visualize the posterior distributions. We test the $f(T, T_\mathcal{G})$ model using three complementary late-time datasets: (1) Cosmic Chronometers (CC), which provide model-independent measurements of the Hubble parameter $H(z)$ through the differential age evolution of elliptical galaxies~\cite{Moresco_2022_25}; (2) the Pantheon$^+$ samples without the SH0ES calibration (PPS), comprising 1701 SNe Ia extending up to redshift $0 \leq z\leq 2.3$~\cite{Brout_2022_938, Scolnic_2022_938}; and (3) BAO$_1$, DESI BAO DR1 and DESI BAO DR2, which offer precise distance measurements from galaxy clustering and Lyman-$\alpha$ forest observations over the redshift range $0.3 \leq z \leq 2.33$ \cite{Adame_2024_DESI_collaboration, DESI_DR2}. This multi-probe approach combines independent distance ladders (SNe~Ia), expansion rate measurements (CC), and large-scale structure constraints (BAO), providing complementary constraints with reduced dependence on a single baseline cosmological model.~\cite{Agostino_2018_98, Agostino_2019_99, Lohakare_2023_40_CQG}. The subsequent sections detail the distinctive features of each dataset and their corresponding likelihood functions, with particular emphasis on the improved precision offered by \textsc{DESI} DR2's anisotropic clustering measurements compared to earlier BAO surveys.

    Before discussing the datasets in detail, we stress that the present study is intentionally restricted to late-time probes. Our aim is to test whether the model can reproduce the observed low-redshift expansion history and whether it can shift the inferred value of $H_0$ relative to a simpler late-time baseline. A full confrontation with CMB TT/TE/EE likelihoods or BBN constraints would require extending the analysis to a complete early-Universe treatment, including recombination-era perturbation transfer and sound-horizon physics. That broader program is beyond the scope of the present work and is therefore left for future study.
    
\subsection{Cosmological datasets}
\subsubsection{Cosmic Chronometers (CC)} \label{SEC III a}
    The CC method provides a model-independent measurement of the Hubble parameter $H(z)$ by estimating the differential age evolution $dt$ of passively evolving elliptical galaxies across small redshift intervals ($ \delta z/z \leq 0.001 $) \cite{Jimenez_2002_573}. This technique relies on high-precision spectroscopy of these ``chronometers,'' which formed simultaneously but are observed at slightly different redshifts, enabling a direct reconstruction of the expansion history of the Universe without cosmological assumptions. For our analysis, we employ the most recent CC dataset \cite{Moresco_2022_25, Lohakare_2023_40_CQG}, which incorporates a comprehensive covariance matrix to account for systematic uncertainties from stellar population synthesis models, including the initial mass function, stellar library, and metallicity dependencies. This approach builds upon earlier foundational work \cite{Moresco_2012_2012_006, Moresco_2015_450, Moresco_2016_2016_014} while rigorously addressing both statistical and systematic errors to deliver robust $H(z)$ constraints.

\subsubsection{Supernovae type Ia (SNe Ia)} \label{SEC III b}
    SNe Ia are extremely bright stellar explosions that can be detected across the Universe, even at redshifts as high as $z \sim 2.3$. They are widely regarded as premier cosmological distance indicators due to the remarkable homogeneity of their spectra and light curves, combined with their widespread distribution across the sky. The distance modulus $\mu$ can be determined from SNe Ia observations using the relation
        \begin{eqnarray}
            \mu = m_b - M\, ,
        \end{eqnarray}
    where $m_b$ is the apparent magnitude (serving as an overall flux normalization) and $M$ is the absolute magnitude. Alternatively, $\mu$ can be expressed theoretically as
        \begin{eqnarray}
            \mu = 25 + 5 \log_{10}(d_L(z)) \, ,    
        \end{eqnarray}
    where $d_L(z)$ is the luminosity distance, defined as
        \begin{eqnarray}
            d_L(z) = (1 + z) \int_0^z \frac{c}{H(z')} \, dz' \, ,        
        \end{eqnarray}
    and $c$ is the speed of light in vacuum. The inclusion of PPS data, calibrated by SH0ES Cepheid observations, eliminates the need for an independent absolute magnitude calibration \cite{Riess_2022_934, Chantada_2023_107, Camarena_2020_2, Conley_2011_192}, allowing $M$ to be treated as a free parameter. While recent studies have questioned aspects of these calibrations \cite{Perivolaropoulos_2024_110, Freedman_2024_985}, the methodology remains valid for cosmological constraints, as evidenced by robust weak lensing results \cite{Abbott_2022_105, Asgari_2021_645}, BAO measurements \cite{Adame_2024_DESI_collaboration}, and SNe Ia distance calibrations \cite{Riess_2022_934}.

\subsubsection{Baryon Acoustic Oscillation (BAO)} \label{SEC III c}
    BAO correspond to density fluctuations in baryonic matter, originating from acoustic density waves in the primordial plasma of the early Universe. The relevant measurements are documented in Ref.~\cite{Adame_2024_DESI_collaboration}, which provides data and correlations for the comoving distance during the drag epoch, expressed as $D_M / r_d$, and the distance variable $D_H / r_d$. Here, $r_d$ denotes the comoving sound horizon at the drag epoch, representing the maximum distance sound waves could propagate from the Big Bang until baryon decoupling. In cases with a low signal-to-noise ratio, the averaged quantity $D_V / r_d$ is employed instead. It should be emphasized that when utilizing DESI BAO data alone, only the combined parameter $r_d H_0$ can be constrained. However, by combining DESI BAO measurements with complementary observational datasets, it becomes possible to independently determine $r_d$ and the Hubble constant $H_0$. For clarity, throughout this paper the label ``BAO'' refers to the pre-DESI compilation denoted by BAO$_1$, while ``DESI 2024'' refers specifically to DESI DR1. DESI DR2 is treated separately as the most recent BAO dataset considered here. Because BAO$_1$ and DESI partially overlap in redshift range and survey volume, they are not combined in a single statistical fit.

    The BAO$_1$ observables listed in Table \ref{Table:BAO1} and the statistics for the DESI samples used in the DESI DR1 and DR2 measurements as presented in Table \ref{Table:BAO stat DR1} and \ref{Table:BAO stat DR2}, respectively (see \hyperref[Appendix: A]{Appendix B}), are defined as follows
    \begin{enumerate}    
    \item Hubble distance $D_H(z)$,
        \begin{equation}
            D_H(z)=\frac{c}{H(z)} \, .
        \end{equation}   
    \item Angular diameter distance $D_A(z)$,
        \begin{equation}
            D_A(z) = \frac{1}{(1+z)} \int_0^z \frac{c}{H(z')}dz' \, .
        \end{equation}
    \item Transverse comoving distance $D_M(z)$,
        \begin{equation}
            D_M(z) = (1+z)D_A(z) \, .
        \end{equation}
    \item Volume-averaged distance $D_V(z)$,
        \begin{equation}
            D_V(z) = \left[ (1+z)^2 D_A^2(z) \frac{c z}{H(z)}  \right]^{1/3} \, .
        \end{equation}
    \end{enumerate}

    We analyze three distinct BAO datasets from various surveys: BAO$_1$ (see Table \ref{Table:BAO1} in \hyperref[Appendix: A]{Appendix B}), comprising 20 data points (from SDSS \cite{Ross_2015_449, Bautista_2017_603, Bautista_2020_500, Neveux_2020_499, Ata_2017_473, Alam_2017_470}, DES \cite{Abbott_2018_483}, and the WiggleZ Dark Energy Survey \cite{Kazin_2014_441}), DESI DR1, consisting of 7 measurements from the DESI 2024 survey \cite{Adame_2024_DESI_collaboration} and DESI DR2 2025 release \cite{DESI_DR2}. Due to partial overlaps in sky regions and redshift ranges between DESI and prior surveys in BAO$_1$, these datasets should not be combined in a single statistical analysis, as correlations among shared objects, particularly at low redshifts, arise from differences in instrument performance and observing strategies, despite some agreement and ongoing discussions regarding inconsistencies, with DESI data considered more accurate due to enhanced fitting and data processing techniques. The dataset includes two isotropic BAO measurements: the Bright Galaxy Survey (BGS) at an effective redshift $z_{\mathrm{eff}} = 0.30$ and quasars (QSO) at $z_{\mathrm{eff}} = 1.49$, along with ten anisotropic BAO data points from the Luminous Red Galaxy (LRG) sample at $z_{\mathrm{eff}} = 0.51$ and $0.71$, LRG+Emission Line Galaxy (ELG) at $z_{\mathrm{eff}} = 0.93$, ELG alone at $z_{\mathrm{eff}} = 1.32$, and Lyman-$\alpha$ quasars (Lya QSOs) at $z_{\mathrm{eff}} = 2.33$. BAO, a powerful cosmological distance estimator, originates from pre-recombination density perturbations in the baryon-photon fluid, where gravitational potential wells drove acoustic waves that imprinted a characteristic scale in the large-scale structure. For enhanced precision, we also incorporate data from the DESI DR2 release \cite{DESI_DR2}, which offers improved measurements due to an expanded galaxy and quasar sample compared to DR1 \cite{Adame_2024_DESI_collaboration}. A key advancement in DR2 is the ability to independently constrain the transverse comoving distance $D_M(z)$ and Hubble distance $D_H(z)$, whereas DR1 provided only the volume-averaged distance $D_V(z)$, highlighting the significance of the higher signal-to-noise ratio in the latest quasar dataset.

    For the $f(T, T_{\mathcal{G}})$ gravity model investigated in this manuscript, we adopt the following prior ranges for the model parameters: $H_0 \in [50.0, 100.0] \, \mathrm{km} \, \mathrm{s}^{-1} \, \mathrm{Mpc}^{-1}$, $\Omega_{\mathrm{m}0} \in [0.01, 1.0]$, $\alpha \in [-5.0, 5.0]$, $\beta \in [-5.0, 5.0]$, $\gamma \in [0.0, 50.0]$, and $M \in [-20.0, -18.0]$. These priors are used in the MCMC analysis to constrain the model against observational datasets.
    \begin{table*}
        \centering
    \resizebox{1\textwidth}{!}{
    \begin{tabular}{|*{7}{c|}}\hline
        {\centering  \textbf{Datasets}} & $H_0$ & $\Omega_{\mathrm{m}0}$ & $\alpha$ & $\beta$ & $\gamma$ & $M$ \\ [0.5ex]
    \hline \hline
        \parbox[c][0.8cm]{3cm}{\centering \textbf{CC+PPS}} & $71.411^{+1.214}_{-1.231}$ & $0.291^{+0.021}_{-0.023}$ & $1.301^{+0.051}_{-0.052}$ & $-2.507^{+0.181}_{-0.180}$ & $31.170^{+0.132}_{-0.132}$ & 
        $-19.291^{+0.080}_{-0.070}$ \\
    \hline
        \parbox[c][0.8cm]{3cm}{\centering \textbf{CC+PPS+BAO$_1$}} & $70.168^{+0.850}_{-0.900}$ & $0.322^{+0.018}_{-0.017}$ & $1.211^{+0.042}_{-0.041}$ & $-2.301^{+0.120}_{-0.130}$ & $33.891^{+0.110}_{-0.120}$ &
        $-19.351^{+0.070}_{-0.060}$ \\[0.5ex] 
    \hline
        \parbox[c][0.9cm]{3.2cm}{\centering \textbf{CC+PPS+DESI DR1}} & $70.456^{+1.000}_{-1.100}$ & $0.316^{+0.020}_{-0.019}$ & $1.257^{+0.045}_{-0.044}$ & $-2.460^{+0.140}_{-0.150}$ & $33.396^{+0.120}_{-0.125}$ & 
        $-19.224^{+0.060}_{-0.060}$ \\[0.5ex] 
    \hline
        \parbox[c][0.9cm]{3.2cm}{\centering \textbf{CC+PPS+DESI DR2}} & $69.144^{+0.450}_{-0.547}$ & $0.307^{+0.021}_{-0.020}$ & $1.189^{+0.038}_{-0.037}$ & $-2.184^{+0.140}_{-0.130}$ & $31.840^{+0.130}_{-0.120}$ & 
        $-19.378^{+0.070}_{-0.080}$ \\[0.5ex] 
    \hline
    \end{tabular}}
        \caption{Summary of MCMC-constrained cosmological parameters for the data combinations shown (CC+PPS, CC+PPS+BAO$_1$, CC+PPS+DESI DR1, and CC+PPS+DESI DR2). The table presents the posterior estimates for the Hubble constant ($H_0$), matter density parameter ($\Omega_{\mathrm{m}0}$), and modified gravity parameters ($\alpha$, $\beta$, $\gamma$ and $M$) with their $1\sigma$ confidence intervals.}
    \label{Table: mcmc parametrized values}
    \end{table*}
    
    \begin{figure}[H]
    \centering
        \includegraphics[width=160mm, height=210mm]{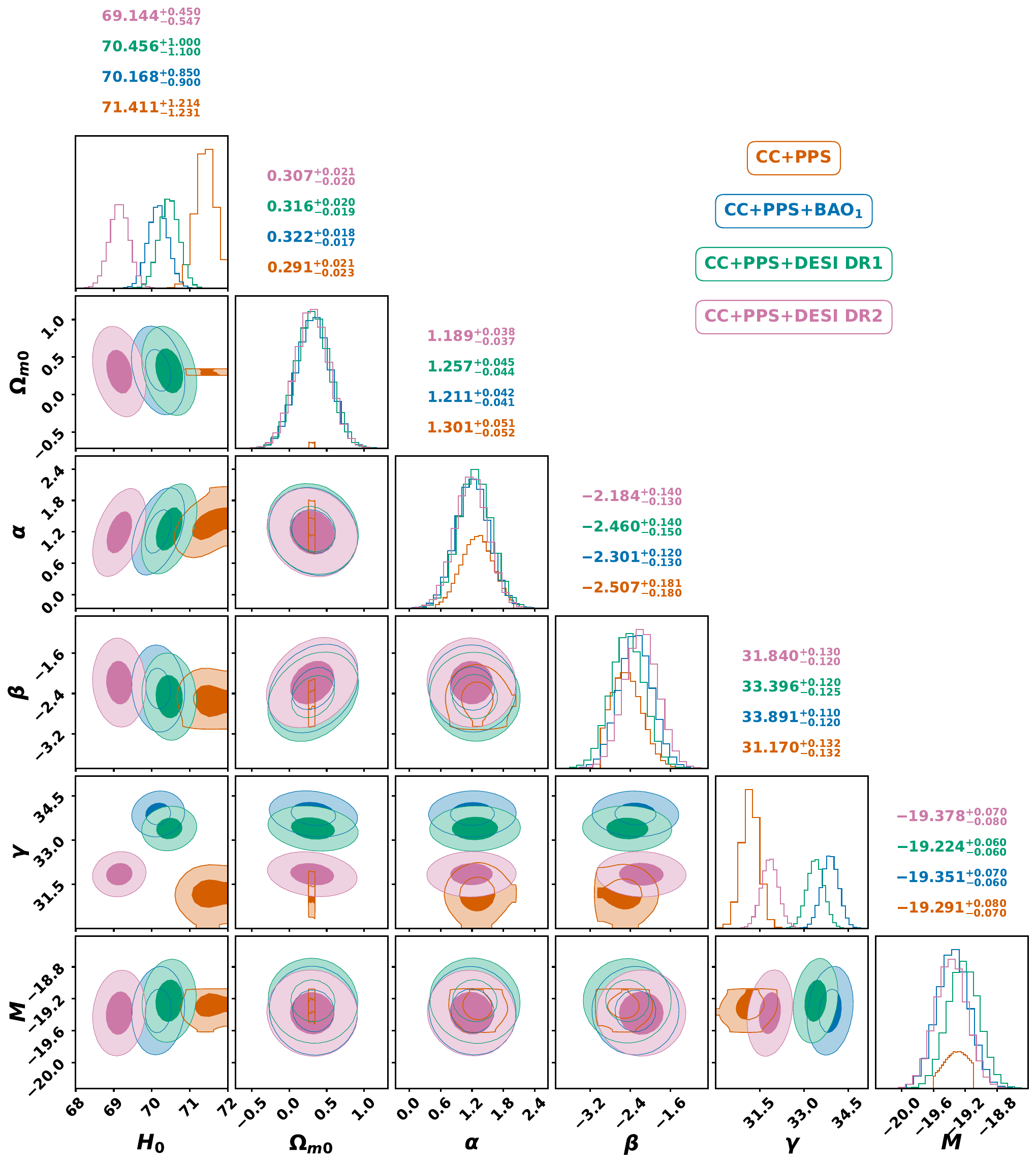}
        \caption{The contour plots display the $1\sigma$ and $2\sigma$ uncertainty regions for the model parameters $H_0$, $\Omega_{\mathrm{m}0}$, $\alpha$, $\beta$, $\gamma$ and $M$. These contours are based on the combined CC+PPS, CC+PPS+BAO$_1$, CC+PPS+DESI DR1 and CC+PPS+DESI DR2 datasets.} 
    \label{FIG: contour}
    \end{figure}

    \begin{figure*}[!htb]
        \centering
        \includegraphics[width=7.7cm, height=6.50cm]{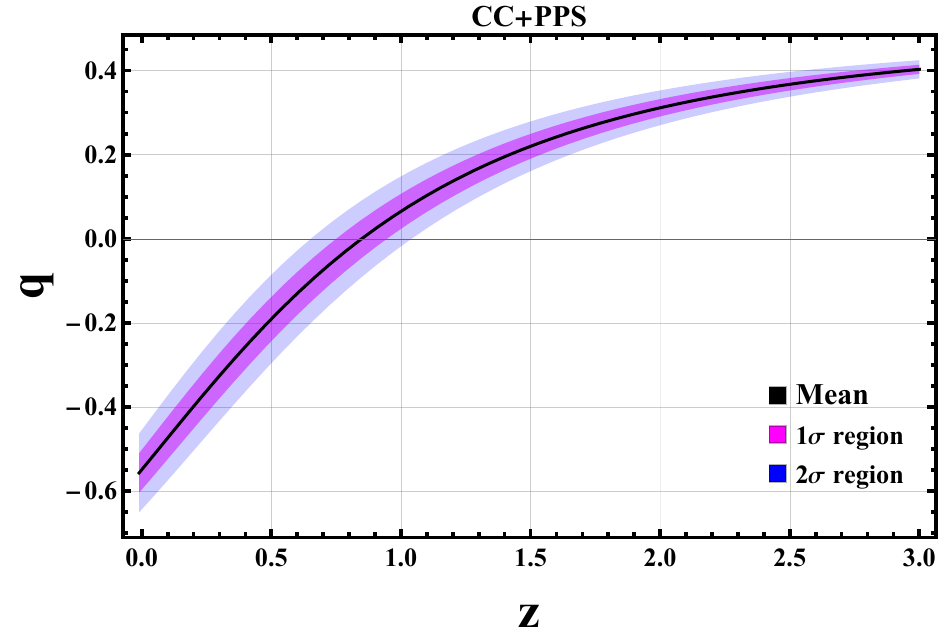}~~~~
        \includegraphics[width=7.7cm, height=6.50cm]{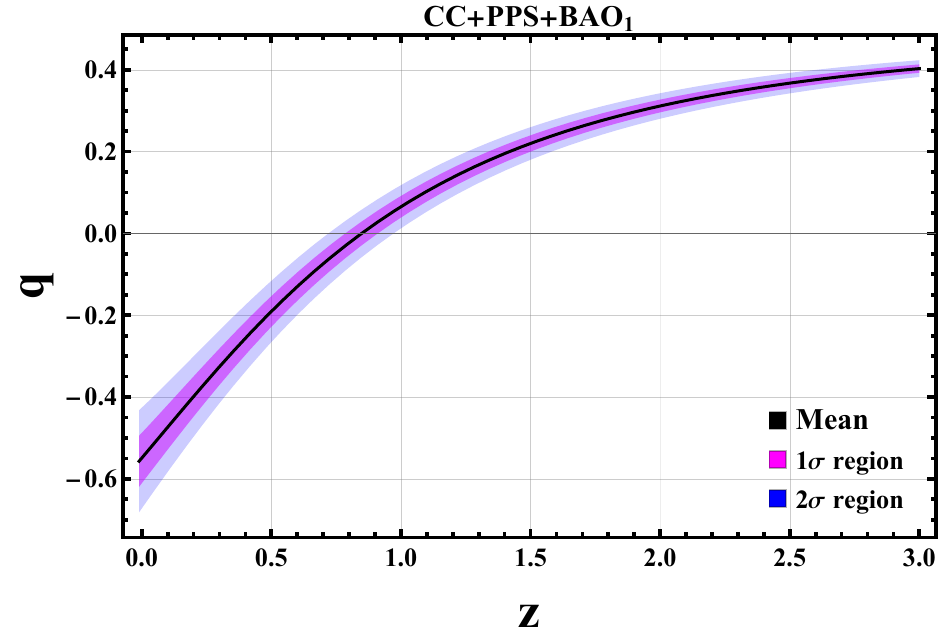}\\
        \includegraphics[width=7.7cm, height=6.5cm]{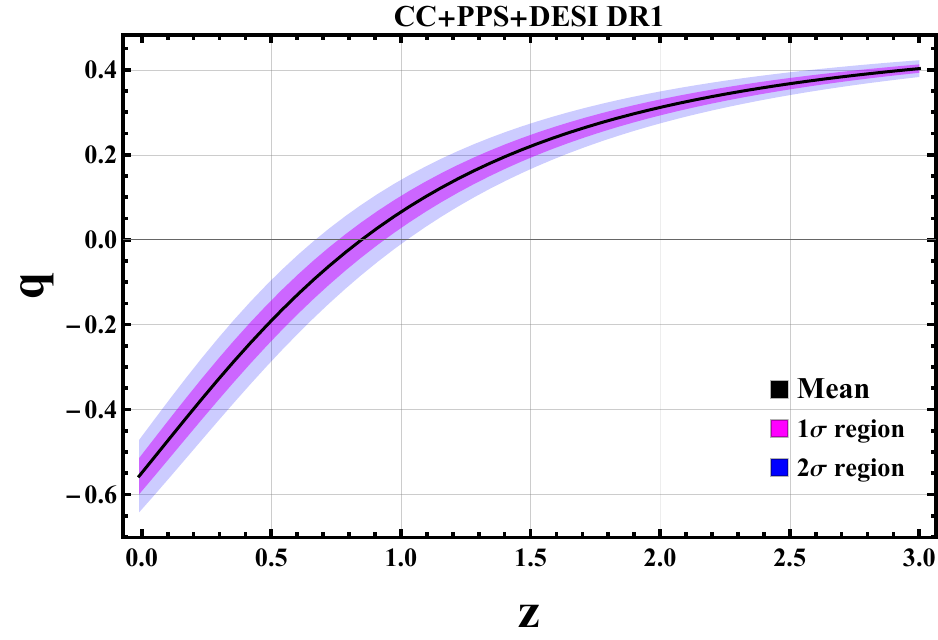}~~~~
        \includegraphics[width=7.7cm, height=6.5cm]{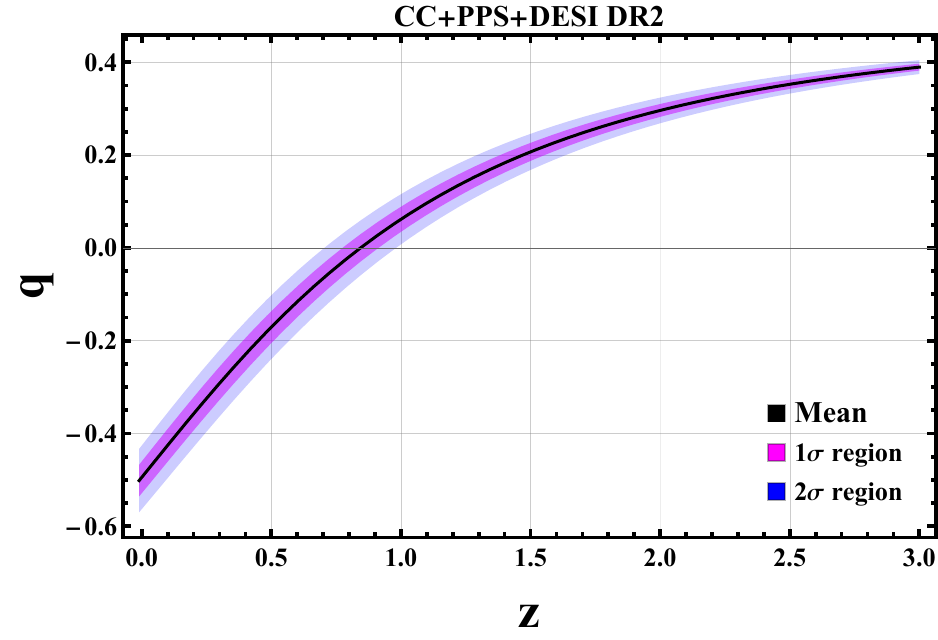}
        \caption{Behavior of the deceleration parameter using the combined datasets.}
        \label{FIG: q}
    \end{figure*}

    Figure \ref{FIG: contour} presents the $1\sigma$ ($68\%$ confidence) and $2\sigma$ ($95\%$ confidence) uncertainty regions for the parameters $H_0$, $\Omega_{\mathrm{m}0}$, $\alpha$, $\beta$, $\gamma$ and $M$, obtained through MCMC analysis. The contours compare constraints from four key datasets: CC+PPS, CC+PPS+BAO$_1$, CC+PPS+DESI DR1 and CC+PPS+DESI DR2 sample. Each pair of parameters shows marginalized posterior distributions, with the inner and outer contours delineating progressively tighter and broader confidence intervals, respectively. Figure \ref{FIG: contour} presents the constraints on the Hubble parameter $H_0$ derived from an MCMC analysis within the $f(T, T_\mathcal{G})$ gravity framework, utilizing various cosmological datasets. The best-fit values of $H_0$ are reported with their $68\%$ confidence intervals, revealing a range of estimates such as $69.144_{-0.547}^{+0.450}$, $70.456_{-1.100}^{+1.000}$, $70.168_{-0.900}^{+0.850}$ and $71.411_{-1.234}^{+1.214}$. These results highlight the consistency of the $f(T, T_\mathcal{G})$ model across different datasets, with $H_0$ values clustering around 69--71.5\kms, aligning closely with recent observational bounds \cite{Escudero_2026_998, Freedman_2024_985, Sharma_2026_034}. The tight constraints, particularly for datasets yielding $H_0 = 69.144_{-0.547}^{+0.450}$, underscore the model's robustness, while the slight variations across datasets reflect the influence of different observational priors. This analysis provides valuable insights into the viability of $f(T, T_\mathcal{G})$ gravity in describing late-time cosmic expansion, offering a pathway to address tensions in $H_0$ measurements.
    
    Figure~\ref{FIG: q} highlights the critical role of the deceleration parameter $q=-1+\frac{(1+z)H(z)H'(z)}{H(z)^2}$, a fundamental cosmological quantity that characterizes the expansion dynamics of the Universe. A positive value of $q$ corresponds to a decelerating expansion, while a negative value indicates an accelerating phase. Analyzing the CC+PPS, CC+PPS+BAO$_1$, CC+PPS+DESI DR1, and CC+PPS+DESI DR2 datasets reveals a consistent evolution of $q$: it transitions from positive in the past signifying early deceleration to negative in the present, reflecting the current accelerated expansion, as shown in Figure~\ref{FIG: q}. The present-day values of the deceleration parameter $q_0$ are found to be $-0.549$, $-0.538$, $-0.549$, and $-0.494$ for the respective datasets. These estimates align well with the observational constraint $q_0 = -0.528^{+0.092}_{-0.088}$ reported by \cite{Christine_2014_89}.

    The model also indicates a smooth transition redshift, $z_t$, where the Universe shifted from deceleration to acceleration. The derived transition redshifts are $z_t = 0.834$, $0.842$, $0.846$, and $0.837$ for the CC+PPS, CC+PPS+BAO$_1$, CC+PPS+DESI DR1, and CC+PPS+DESI DR2 datasets, respectively. These results are consistent with existing observational constraints, including $H(z)$ measurements at $z \approx 2.3$ from BAO data \cite{Busca_2013_552}, and a transition redshift estimated as $z_t = 0.74 \pm 0.5$ \cite{Farooq_2013_766}, $z_t = 0.7679^{+0.1831}_{-0.1829}$ \cite{Capozziello_2014_90_044016}, and $z_t = 0.60^{+0.21}_{-0.12}$ \cite{Yang_2020_2020_059}.
    \begin{figure*}[!htb]
        \centering
        \includegraphics[width=7.7cm, height=6.50cm]{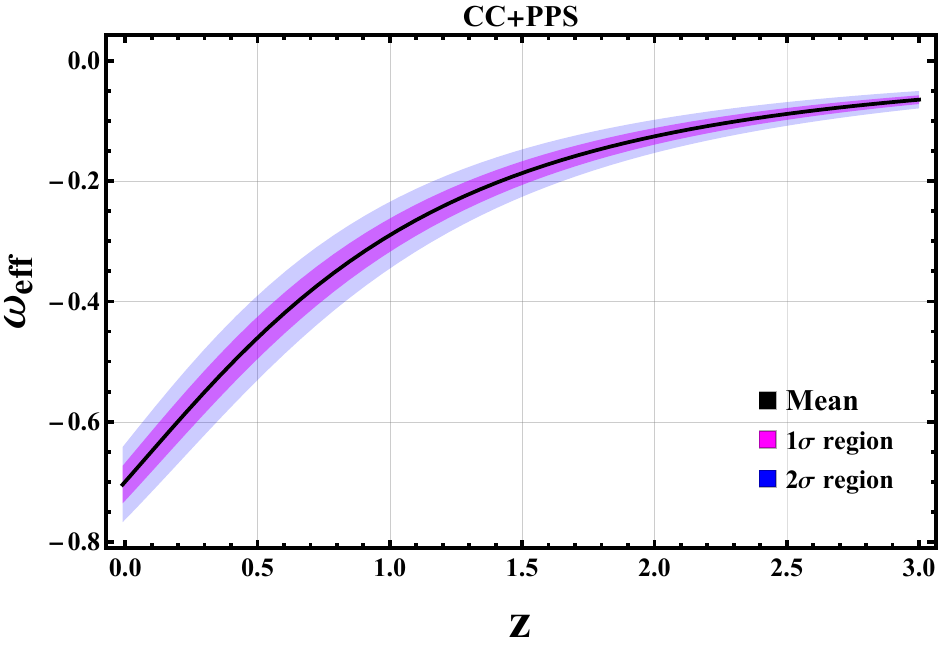}~~~~
        \includegraphics[width=7.7cm, height=6.50cm]{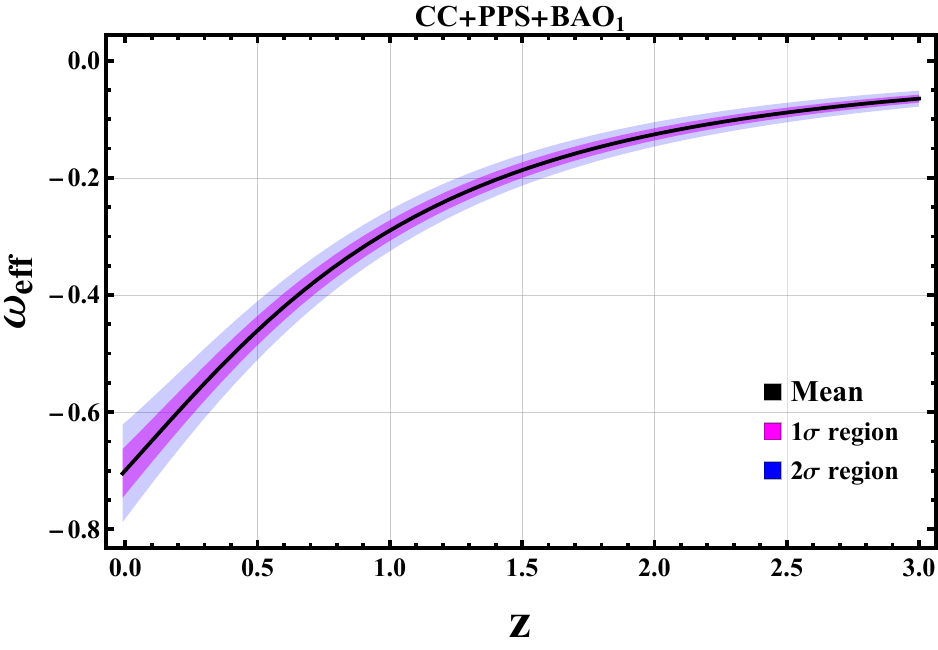}\\
        \includegraphics[width=7.7cm, height=6.50cm]{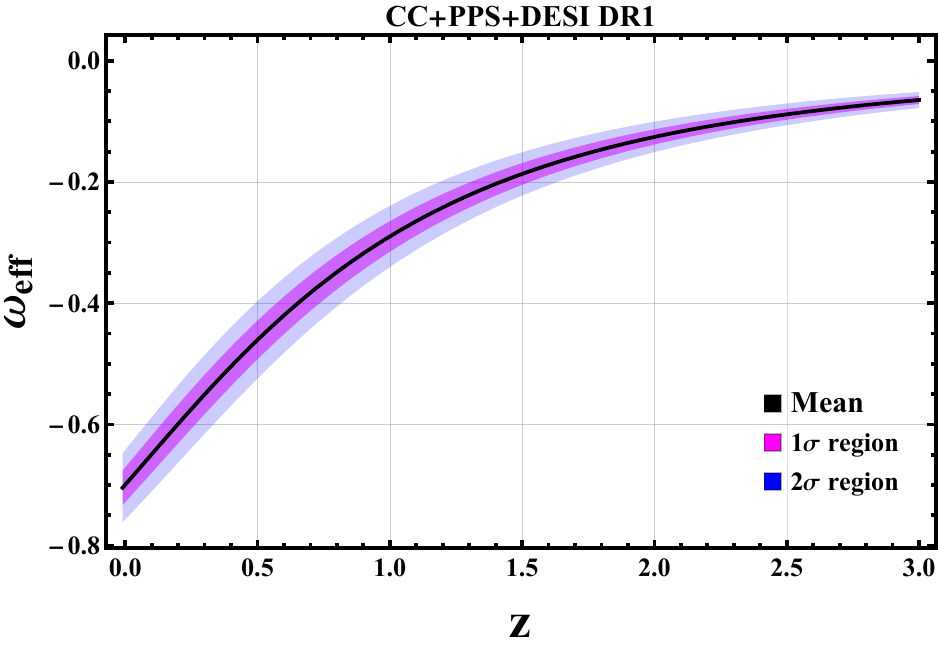}~~~~
        \includegraphics[width=7.7cm, height=6.50cm]{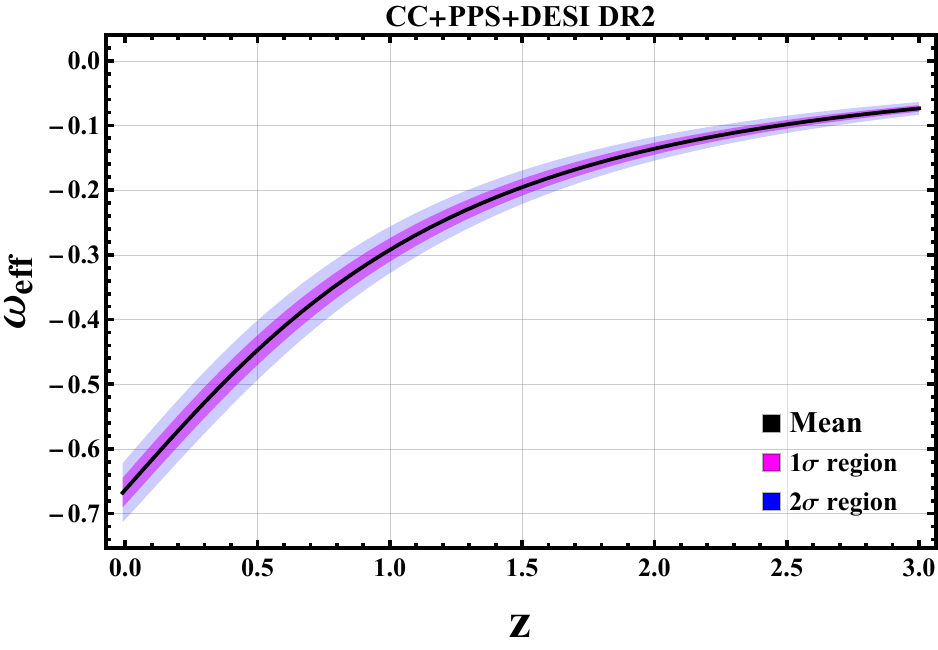}
        \caption{Behavior of the dark energy EoS parameter using the combined datasets.}
        \label{FIG: eos}
    \end{figure*}

    Additionally, effective EoS parameter $\omega_{\mathrm{eff}}=-1+\frac{2(1+z)H(z)H'(z)}{3H(z)^2}$, plays a pivotal role in tracing the nature of the energy components driving cosmic acceleration. Figure~\ref{FIG: eos}, redshift evolution of the effective EoS parameter $\omega_{\text{eff}}$ for the $f(T, T_\mathcal{G})$ gravity model, derived from the CC+PPS, CC+PPS+BAO$_1$, CC+PPS+DESI DR1, and CC+PPS+DESI DR2 datasets. The present-day values range from $-0.664$ to $-0.693$, consistent with a quintessence-like dark energy component. The slight steepening at low redshifts suggests a dynamic dark energy behavior, remaining above the phantom regime ($\omega_{\mathrm{eff}} < -1$). These are consistent with previous constraints such as those from Planck 2018 \cite{Aghanim_2020_641} and from WMAP+CMB analyses \cite{Hinshaw_2013_208}. The evolution of the effective EoS, derived from the computed energy density and pressure of dark energy, is also presented as a function of redshift in Figure~\ref{FIG: eos}.

\subsection{Statistical analysis}
    It is important to clarify the meaning of the $\Lambda$CDM comparison used in this work. The baseline model considered here is not the full six-parameter CMB $\Lambda$CDM framework, but a reduced late-time background implementation appropriate to the observables analyzed in this paper, namely CC, Pantheon$^+$, and BAO distance measurements. In this setting, the effective free parameters are those that directly govern the low-redshift expansion history together with the relevant nuisance contribution from the supernova sector. The comparison is therefore performed at the same phenomenological level as the modified-gravity model, rather than against a Planck-style global fit. We evaluate the performance of the model against the standard $\Lambda$CDM model using the Akaike Information Criterion (AIC) \cite{Akaike_1974_19}, the Bayesian Information Criterion (BIC) \cite{Schwarz_1978_6}, and the minimum chi-squared statistic ($\chi^2_{\mathrm{min}}$). Both AIC and BIC balance the goodness of fit of a model with its complexity, determined by the number of parameters ($n$). The AIC is calculated as
        \begin{equation}
            \mathrm{AIC} = \chi^2_{\mathrm{min}} + 2 n,
        \end{equation}
   where $\chi^2$ is derived from the value of the model's Gaussian likelihood function $\mathcal{L}(\hat{\theta} \mid \text{data})$, evaluated at the best-fit parameters. A lower AIC value indicates a better fit to the data, with a penalty applied to models with more parameters to account for overfitting. Similarly, the BIC is defined as
        \begin{equation}
            \mathrm{BIC} = \chi^2_{\mathrm{min}} + n \ln \mathcal{N},
        \end{equation}
    where $\mathcal{N}$ is the number of data samples used in the MCMC process.

    To compare the $f(T, T_\mathcal{G})$ model with the $\Lambda$CDM benchmark, we compute the differences in AIC and BIC, expressed as
        \begin{equation}
            \Delta \mathrm{AIC} = \Delta \chi^2_{\mathrm{min}} + 2 \Delta n, \quad \Delta \mathrm{BIC} = \Delta \chi^2_{\mathrm{min}} + \Delta n \ln \mathcal{N}.
        \end{equation}
    These metrics quantify deviations from the benchmark, with smaller $\Delta \mathrm{AIC}$ and $\Delta \mathrm{BIC}$ values indicating closer alignment with the $\Lambda$CDM model, suggesting superior model performance relative to the dataset. This comparative analysis provides deeper insights into the consistency of each model with the standard cosmological framework.
   
\begin{table*}
    \centering
    \resizebox{1\textwidth}{!}{
    \begin{tabular}{|c|c|c|c|c|c|c|c|c|c|}
    \hline
        {Data sets}  &\multicolumn{2}{c|}{$\chi^2_{\text{min}}$} 
                     &\multicolumn{2}{c|}{AIC}  
                     &\multicolumn{2}{c|}{BIC} 
                     & {$\Delta \text{AIC}$} 
                     & {$\Delta \text{BIC}$} 
                     & {$\mathcal{N}$} \\
    \cline{2-7}   
        & $f(T,T_\mathcal{G})$ & $\Lambda$CDM 
        & $f(T,T_\mathcal{G})$ & $\Lambda$CDM 
        & $f(T,T_\mathcal{G})$ & $\Lambda$CDM 
        & & & \\ 
    \hline \hline
        CC+PPS & 1661.412 & 1682.482 & 1671.412 & 1688.482 & 1698.700 & 1704.855 & -17.070 & -6.155 & 32+1701 \\
    \hline
        CC+PPS+BAO$_1$ & 1676.329 & 1691.321 & 1686.329 & 1697.321 & 1713.674 & 1713.728 & -10.992 & -0.054 & 32+1701+20 \\
    \hline
        CC+PPS+DESI DR1 & 1670.364 & 1688.234 & 1683.253 & 1694.234 & 1710.564 & 1710.619 & -10.981 & -0.055 & 32+1701+7 \\
    \hline
        CC+PPS+DESI DR2  & 1671.238 & 1685.345 & 1678.349 & 1691.345 & 1705.660 & 1707.733 & -12.996 & -2.073 & 32+1701+9 \\
    \hline
    \end{tabular}}
    \caption{Goodness-of-fit statistics for the $f(T, T_\mathcal{G})$ gravity model: Minimum $\chi^2$ values with corresponding $\Delta \mathrm{AIC}$ and $\Delta \mathrm{BIC}$ measures. The lower section compares this information criterion against the $\Lambda$CDM reference model, highlighting the relative performance of both cosmological scenarios.}
    \label{Table: statistical values}
\end{table*}

    The $\Lambda$CDM comparison in this work refers to a reduced late-time background implementation appropriate to the CC, Pantheon$^+$, and BAO likelihoods, not to the full six-parameter CMB $\Lambda$CDM model. In this setting, the effective free parameters are those directly entering the low-redshift observables. We report the minimum $\chi^2$ to quantify fit quality and use AIC/BIC to assess whether any improvement justifies the additional freedom of the modified-gravity model. Table~\ref{Table: statistical values} presents the goodness-of-fit statistics for the $f(T,T_\mathcal{G})$ gravity model, evaluated against a reduced late-time $\Lambda$CDM baseline constructed for the same CC, Pantheon$^+$, and BAO likelihoods used in this analysis. For $\Lambda$CDM, the minimum chi-squared ($\chi^2_{\mathrm{min}}$), AIC, and BIC values across the datasets are as follows: for CC+PPS, $\chi^2_{\mathrm{min}} = 1682.482$, AIC = 1688.482, BIC = 1704.855; for CC+PPS+BAO$_1$, $\chi^2_{\mathrm{min}} = 1691.321$, AIC = 1697.321, BIC = 1713.728; for CC+PPS+DESI DR1, $\chi^2_{\mathrm{min}} = 1688.234$, AIC = 1694.234, BIC = 1710.619; and for CC+PPS+DESI DR2, $\chi^2_{\mathrm{min}} = 1685.345$, AIC = 1691.345, BIC = 1707.733. These values serve as the reference for computing $\Delta \mathrm{AIC}$ and $\Delta \mathrm{BIC}$ for the $f(T, T_{\mathcal{G}})$ model. For the CC+PPS dataset, the $f(T, T_{\mathcal{G}})$ model yields $\chi^2_{\mathrm{min}} = 1661.412$, with $\Delta \mathrm{AIC} = -17.070$ and $\Delta \mathrm{BIC} = -6.155$, indicating a strong statistical preference over $\Lambda$CDM due to the significant negative values of both criteria. This suggests that the $f(T, T_{\mathcal{G}})$ model provides a better fit to the data despite its additional complexity. Similarly, for CC+PPS+DESI DR2, the $f(T, T_{\mathcal{G}})$ model achieves $\chi^2_{\mathrm{min}} = 1671.238$, with $\Delta \mathrm{AIC} = -12.996$ and $\Delta \mathrm{BIC} = -2.073$, again suggesting a moderate to strong preference over $\Lambda$CDM. For CC+PPS+BAO$_1$ and CC+PPS+DESI DR1, the $f(T, T_{\mathcal{G}})$ model records $\chi^2_{\mathrm{min}}$ values of 1676.329 and 1670.364, respectively, with $\Delta \mathrm{AIC}$ values of -10.992 and -10.981, and $\Delta \mathrm{BIC}$ values of -0.054 and -0.055. These results indicate a moderate preference for the $f(T, T_{\mathcal{G}})$ model based on AIC, while BIC shows a marginal difference, reflecting the penalty for additional free parameters in the modified gravity model compared to three parameters of the $\Lambda$CDM. Overall, the consistently negative $\Delta \mathrm{AIC}$ and $\Delta \mathrm{BIC}$ values across all datasets highlight the competitive performance of the $f(T, T_{\mathcal{G}})$ model, positioning it as a viable alternative to $\Lambda$CDM. However, the near-zero $\Delta \mathrm{BIC}$ values for CC+PPS+BAO$_1$ and CC+PPS+DESI DR1 suggest that the additional parameters in the $f(T, T_{\mathcal{G}})$ model may not be strongly justified by the data under the BIC criterion, underscoring the need for further refinement to balance model complexity and fit.    
    \begin{figure*} [!htb]
        \centering
        \includegraphics[width=155mm,height=165mm]{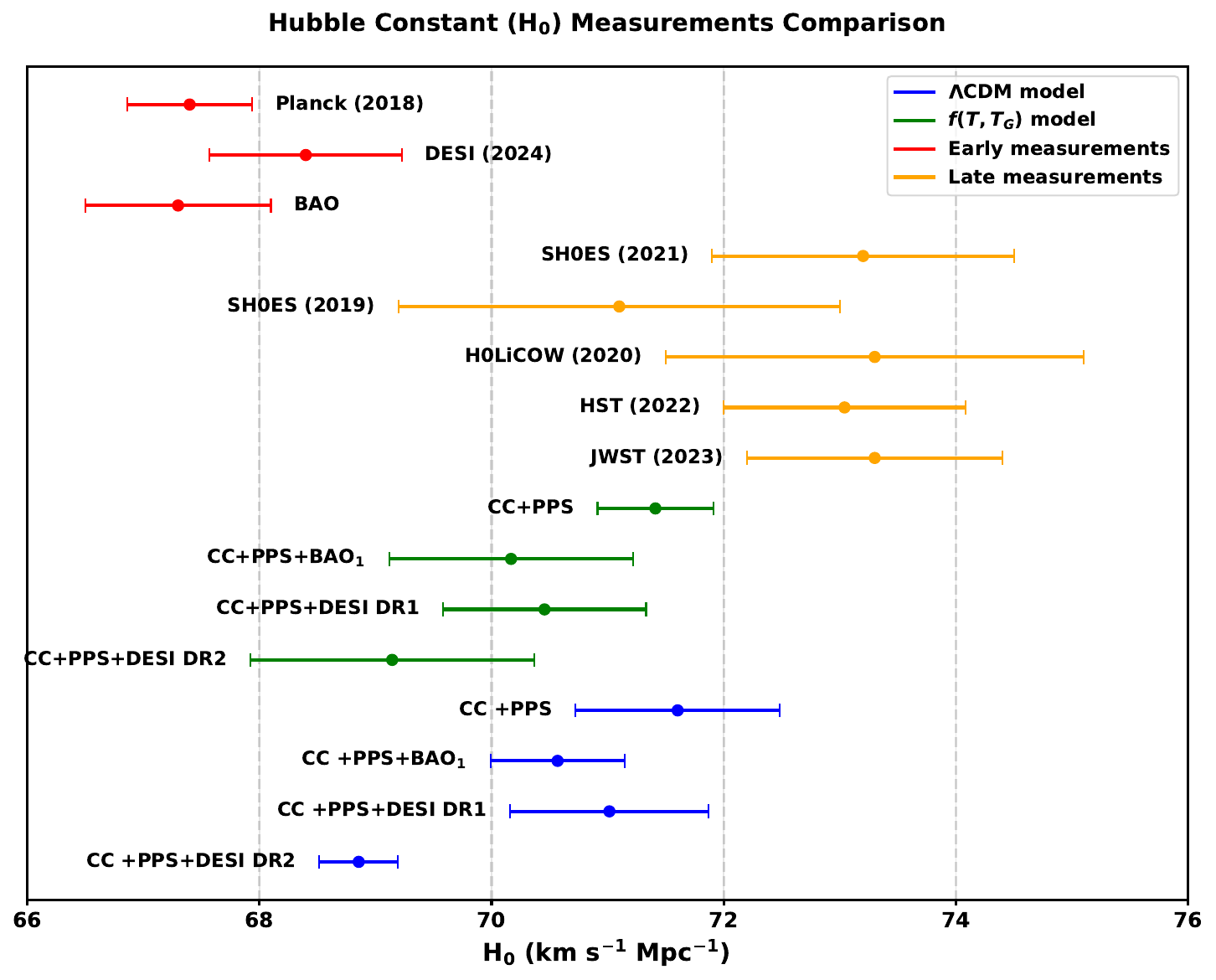}
        \caption{Comparison of the observational BAO distance measurements with the theoretical predictions of the $f(T,T_\mathcal{G})$ model and the corresponding late-time $\Lambda$CDM baseline. Here, ``BAO'' denotes the pre-DESI BAO$_1$ compilation (including SDSS, DES, WiggleZ, and related surveys), while ``DESI 2024'' denotes DESI DR1. DESI DR2 is shown separately as the most recent dataset considered in this work. The plotted $f(T,T_\mathcal{G})$ and $\Lambda$CDM curves are evaluated using the corresponding best-fit or posterior-supported parameter values obtained within the same late-time likelihood framework adopted in this paper; no external CMB prior is imposed in this figure.}
        \label{FIG: whisker_plot}
    \end{figure*}
    
    In Figure \ref{FIG: whisker_plot}, we compare the $H_0$ values derived from joint analyses (as detailed in Table \ref{Table: mcmc parametrized values}) for both the $\Lambda$CDM and $f(T, T_\mathcal{G})$-gravity models against a range of contemporary measurements. The posterior distributions also reveal nontrivial correlations among $H_0$, $\Omega_{m0}$, and the model parameters $(\alpha,\beta,\gamma)$, as expected in an extended late-time expansion history. These degeneracies are not, by themselves, pathological; rather, they reflect the fact that several parameter combinations can reproduce similar low-redshift distance behavior. What matters physically is that the shift toward larger $H_0$ values is not confined to a single isolated point in parameter space, but appears across a broader posterior-supported region. These comparisons are categorized into: i) early-Universe observations, including Planck 2018 and the DESI collaboration; and ii) late-time measurements from direct observations of the local Universe, such as Supernovae and $H_0$ for the Equation of State (SH0ES) \cite{Riess_2022_934}, H0LiCOW \cite{Wong_2019_498}, and the Hubble Space Telescope (HST) \cite{Chen_2016_462}. Figure \ref{FIG: whisker_plot} highlights the discrepancies in $H_0$ measurements between early-Universe data, late-Universe observations, and the theoretical frameworks of $\Lambda$CDM and $f(T, T_\mathcal{G})$-gravity explored in this study. The standard $\Lambda$CDM model aligns closely with early-Universe measurements, which typically predict a lower $H_0$, whereas late-time observations from sources like SH0ES and JWST consistently report higher $H_0$ values. In contrast, the $f(T, T_\mathcal{G})$ gravity model produces elevated $H_0$ estimates, indicating a closer alignment with late-time observations, though it does not fully reconcile the ongoing tension in $H_0$ measurements. The total uncertainty in the Hubble parameter computed, represented in Fig. \ref{FIG: whisker_plot} as $\sigma_{\mathrm{total}} = \sqrt{\sigma_{\mathrm{stat}}^2 + \sigma_{\mathrm{sys}}^2}$. This value integrates the statistical uncertainty $\sigma_{\mathrm{stat}}$ derived from MCMC sampling and the systematic uncertainty $\sigma_{\mathrm{sys}}$ associated with observational datasets. The posterior exploration is performed with multiple \texttt{emcee} walkers over the prior ranges listed above, and the resulting chains are analyzed with \texttt{GetDist}. After discarding the burn-in phase, convergence is assessed through chain stabilization and posterior consistency across walkers. The quoted best-fit values are identified from the converged posterior sample as the points yielding the minimum $\chi^2$ in the explored parameter space.

\section{Perturbations and stability} \label{SEC IV}
    In this section, we investigate the stability conditions of cosmological solutions under linear, isotropic, and homogeneous perturbations within the framework of $f(T, T_\mathcal{G})$ gravity. We derive the general perturbation equations for a spatially flat FLRW Universe and apply them to analyze both de Sitter and power-law cosmological solutions.

    We consider a general ansatz for the Hubble parameter, given by
        \begin{equation} \label{hubble sol}
            H(t) = \bar{H}(t)\, ,
        \end{equation}
    which satisfies the modified field equations of motion in $f(T, T_\mathcal{G})$ gravity. Based on this Hubble function, the torsion scalar $T$ and the Gauss-Bonnet invariant $T_\mathcal{G}$ are expressed as
        \begin{equation}
            \bar{T}(t) = 6\bar{H}^2(t)\, , 
        \end{equation}
        \begin{equation}
            \bar{T}_{\mathcal{G}}(t) = 24\bar{H}^2(t) \dot{\bar{H}}(t) + 24\bar{H}^4(t). 
        \end{equation}

    Assuming a specific form of the function $f(T, T_\mathcal{G})$ that yields the cosmological solution given in equation \eqref{hubble sol}, the modified Friedmann equations must satisfy
        \begin{equation}
            -3\bar{H}^2 \left(3\bar{f}_{T_\mathcal{G}} + 2\bar{f}_T\right) + 3\bar{H} \dot{\bar{f}}_{T_\mathcal{G}} - 3\dot{\bar{H}} \bar{f}_{T_\mathcal{G}} + \frac{1}{2}\bar{f} = \kappa^2 \rho_{\mathrm{m}0}\, ,
        \end{equation}
    where $\bar{f}_T$ and $\bar{f}_{T_\mathcal{G}}$ denote partial derivatives of the function $f$ with respect to $T$ and $T_\mathcal{G}$, respectively, evaluated on the background solution.

    Additionally, the conservation equation for the matter energy density $\rho_{\mathrm{m}0}$ with EoS parameter $\omega$ is given by
        \begin{equation} \label{continuity equation}
            \dot{\bar{\rho}}_{\mathrm{m}} + 3\bar{H} \bar{\rho}_{\mathrm{m}} = 0 \, .
        \end{equation}
    This setup establishes the groundwork for studying the dynamical behavior and stability of various cosmological models within the $f(T, T_{\mathcal{G}})$ gravity paradigm.
    
    We define the perturbation for the Hubble parameter and energy density as follows
        \begin{equation} \label{perturb eqn}
            H(t)=\bar{H}(t)\left(1+\delta (t)\right), \hspace{0.5cm} \rho_{\mathrm{m}}(t)=\bar{\rho}_{\mathrm{m}} \left(1+\delta_{\mathrm{m}}(t)\right),
        \end{equation}
    where $\delta(t)$ and $\delta_\mathrm{m}(t)$ denote the dimensionless perturbations associated with the Hubble rate and the matter overdensity, respectively.

    To analyze the dynamics of these perturbations within the linear regime, we perform a Taylor series expansion of the function $f(T, T_\mathcal{G})$ about the background values of the torsion scalar $\bar{T}$ and the Gauss-Bonnet invariant $\bar{T}_{\mathcal{G}}$, corresponding to the unperturbed solution $H(t) = \bar{H}(t)$. The expansion reads
        \begin{equation} \label{general form of model}
            f(T, T_\mathcal{G})=\bar{f} + \bar{f}_T (T-\bar{T})+ \bar{f}_{T_\mathcal{G}} (T_\mathcal{G}-\bar{T}_{\mathcal{G}})+\mathcal{O}^2,
        \end{equation}
    where $\bar{f} = f(\bar{T}, \bar{T}_{\mathcal{G}})$. The term $\mathcal{O}^2$ includes second-order and higher-order contributions in $T$ and $T_\mathcal{G}$, which are formally retained but do not contribute to the linearized perturbation dynamics.

    Substituting the perturbed expressions from equation \eqref{perturb eqn} into the modified FLRW background equation (originally denoted as equation \eqref{Eq: first_field}), and employing the linear expansion from equation \eqref{general form of model}, we obtain the evolution equation governing the perturbation $\delta(t)$ in the linear approximation
        \begin{equation} \label{eq:perturbation_ode}
            c_{2}\ddot{\delta}(t) + c_{1}\dot{\delta}(t) + c_{0}\delta (t)=c_{\mathrm{m}} \delta_{\mathrm{m}}(t) \, ,
        \end{equation}
    where $c_0$, $c_1$, $c_2$, and $c_\mathrm{m}$ (as provided in the \hyperref[Appendix: B]{Appendix A}) are time-dependent coefficients determined by the background solution and the form of $f(T, T_\mathcal{G})$. This formulation allows for a systematic analysis of the stability of the cosmological model under isotropic scalar perturbations in modified gravity theories involving curvature invariants. The matter continuity equation \eqref{continuity equation} yields a second perturbed equation when linearized using the expressions in~\eqref{perturb eqn}
        \begin{equation}
            \dot{\delta}_\mathrm{m}(t) + 3\bar{H}(t)\delta(t) = 0.
            \label{eq:matter_perturbation}
        \end{equation}

    The stability of any FLRW cosmological solution in $f(T, T_\mathcal{G})$ gravity can therefore be investigated through the coupled system of differential equations~\eqref{eq:perturbation_ode} and~\eqref{eq:matter_perturbation}. Given the linear nature of equation~\eqref{eq:perturbation_ode}, its general solution naturally decomposes into two physically distinct components:
        \begin{enumerate}
            \item The homogeneous solution $\delta_{\mathrm{hom}}(t)$, representing perturbations intrinsic to the gravitational sector and determined by the specific form of the $f(T, T_\mathcal{G})$ Lagrangian.
            \item The particular solution $\delta_{\mathrm{part}}(t)$, driven exclusively by the matter density perturbations $\delta_\mathrm{m}(t)$.
        \end{enumerate}

    The complete linear solution consequently takes the form
        \begin{equation}
            \delta(t) = \delta_{\mathrm{hom}}(t) + \delta_{\mathrm{part}}(t). \label{eq:general_solution}
        \end{equation}
    
    Before analyzing specific cosmological scenarios, we begin with some general considerations regarding the stability equations in modified gravity. On the cosmological FLRW background we have $T=6H^2\ge 0$, thus $|T|=T$ and $f(T,T_\mathcal{G})\Big|_{\beta=0}=-(1-\alpha)\,T$. Therefore, the gravitational part of the action is proportional to $(1-\alpha)T$, which is equivalent to the Hilbert--Einstein action up to an overall constant factor. This implies a rescaling of the effective gravitational coupling,
        \begin{eqnarray}
            \kappa_{\rm eff}^2=\frac{\kappa^2}{1-\alpha}\qquad(\alpha\neq 1).
        \end{eqnarray}

    Consequently, in any expression derived from the Friedmann equations one must replace $\kappa^2$ by $\kappa_{\rm eff}^2$ when working on the TEGR branch with $\beta=0$ and $\alpha\neq0$. In particular, the coefficient $c_\mathrm{m}$ in \hyperref[Appendix: B]{Appendix A} (originally written as $c_{\rm m}=\kappa^2\rho_{\rm m}$) becomes
        \begin{eqnarray}
            c_{\rm m}=\kappa_{\rm eff}^2\,\rho_{\rm m}=\frac{\kappa^2}{1-\alpha}\,\rho_{\rm m}.
        \end{eqnarray}

    Note that the case $\alpha\to1$ is singular and corresponds to a vanishing effective gravitational coupling; we therefore assume $\alpha\neq1$ throughout the analysis. In particular, it is instructive to recover the standard GR limit by setting the gravitational Lagrangian to the Hilbert-Einstein form, $f(T, T_\mathcal{G}) = -T$. Under this condition, the perturbation equation \eqref{eq:perturbation_ode} simplifies significantly and reduces to
        \begin{equation}
            -18\bar{H}^2\,\delta(t) = c_\mathrm{m} \delta_\mathrm{m}(t), 
        \end{equation}
    which establishes a purely algebraic relationship between the geometric perturbation $\delta(t)$ and the matter overdensity $\delta_\mathrm{m}(t)$. This implies that within GR, the complete behavior of perturbations around a cosmological solution is fully determined by the matter sector, and conversely, the evolution of matter can be directly inferred from the geometric perturbation.

    By combining the explicit form of the coefficients $c_\mathrm{m}$ (given in the \hyperref[Appendix: B]{Appendix A}) with the fundamental relation \eqref{eq:matter_perturbation}, we can directly establish that
        \begin{equation}
            \delta(t) = -\frac{1}{6} \delta_\mathrm{m}(t) \propto a(t)^{1/2} \, .
        \end{equation}
    
    \begin{figure*}
        \centering
        \includegraphics[width=7.7cm, height=6.0cm]{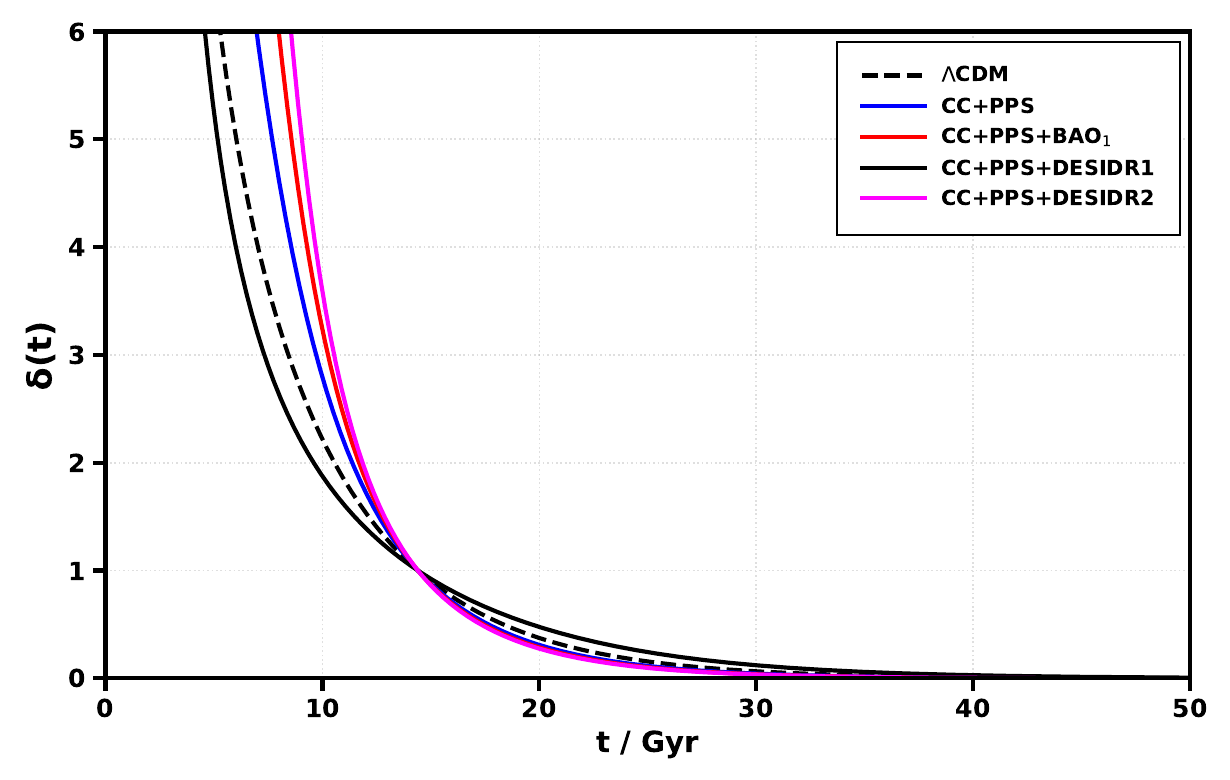}~~~
        \includegraphics[width=7.7cm, height=6.0cm]{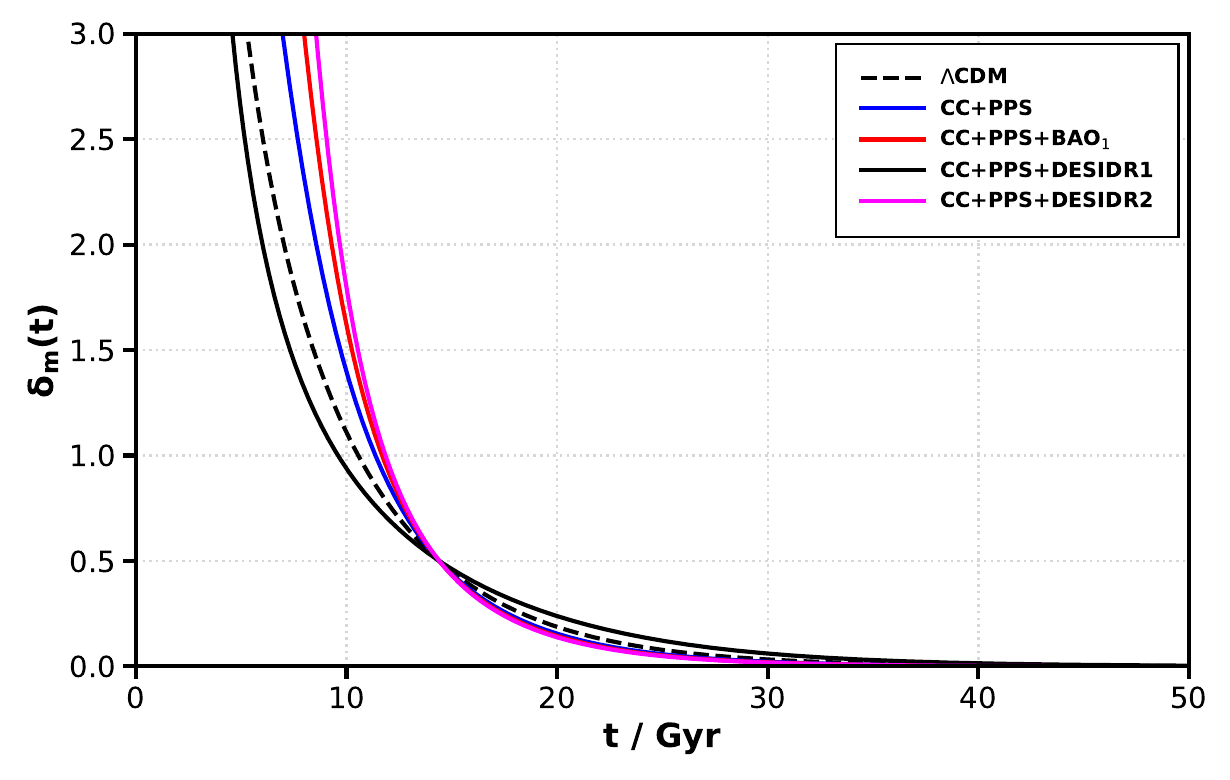}
        \caption{Evolution of the Hubble perturbation parameter $\delta(t)$ (left panel) and the matter perturbation parameter $\delta_{\mathrm{m}}(t)$ (right panel) as a function of cosmic time in the $f(T, T_\mathcal{G})$ gravity model.}
        \label{FIG: delta and delta m}
    \end{figure*}

    However, this algebraic relation is generally absent in higher-order theories of gravity such as those involving arbitrary functions $f(T, T_\mathcal{G})$. In such cases, the evolution of scalar perturbations is governed by the coupled system of differential equations \eqref{eq:perturbation_ode} and \eqref{eq:matter_perturbation}, where the gravitational Lagrangian itself significantly influences the dynamics.
    These coefficients dictate the nature of the perturbation propagation, potentially introducing rich dynamical features absent in standard GR. This foundational contrast underscores the importance of carefully analyzing the stability equations in the context of modified gravity theories, where perturbations exhibit more intricate and model-dependent behavior.

    Figure \ref{FIG: delta and delta m} illustrates the temporal evolution of the Hubble perturbation parameter $\delta(t)$ (left panel) and the matter perturbation parameter $\delta_\mathrm{m}(t)$ (right panel) as functions of cosmic time $t$ in the $f(T, T_\mathcal{G})$ gravity framework. The analysis incorporates the effects of varying the best-fit values of the model parameters, determined through a MCMC parameterization technique. The plots compare four distinct datasets: CC+PPS, CC+PPS+BAO$_1$, CC+PPS+DESI DR1 and CC+PPS+DESI DR2. The numerical results reveal that both $\delta(t)$ and $\delta_\mathrm{m}(t)$ decay rapidly to zero as cosmic time progresses, confirming the stability of the numerical solutions. Fig. \ref{FIG: delta and delta m} illustrates the time evolution of the perturbations. For the Hubble parameter perturbations, the solutions demonstrate stability as they decay monotonically with cosmic time, closely following the behaviour of the $\Lambda$CDM model, and eventually vanish at late times. For matter perturbations, the models again show a behaviour similar to $\Lambda$CDM, remaining stable and decaying with time, but instead of vanishing completely they asymptotically approach a constant limiting value at late times. Specifically, the Hubble perturbations demonstrate a consistent decay across all datasets, fully stabilizing at late times, indicative of a robust and stable cosmological evolution. Similarly, the matter perturbations mirror this behavior, decaying steadily over time and converging to a limiting value at late times, further reinforcing the stability of the solutions within this gravitational model. Within the scalar perturbation sector considered here, and for the posterior-supported region of parameter space explored in the observational analysis, the perturbations remain well-behaved and do not exhibit the instabilities tested in the present framework.

    Building on the stability of the $f(T, T_{\mathcal{G}})$ model confirmed through perturbation analysis, we now summarize its cosmological implications and observational consistency, highlighting its potential as an alternative to $\Lambda$CDM.

\section{Conclusion} \label{SEC conclusion}
    In this study, we investigated the cosmological implications of a specific $f(T, T_{\mathcal{G}})$ gravity model, which incorporates the torsion scalar $T$ and the teleparallel Gauss-Bonnet invariant $T_{\mathcal{G}}$. We began by establishing the theoretical framework of $f(T, T_{\mathcal{G}})$ gravity within a flat FLRW cosmological background, deriving the modified Friedmann equations and exploring the dynamics of cosmic expansion. This model offers a compelling alternative to the standard $\Lambda$CDM paradigm by replicating dark energy behavior without invoking a cosmological constant, thus avoiding associated theoretical challenges such as the fine-tuning and cosmic coincidence problems. We employed a numerical approach to solve the modified Friedmann equations, predicting the redshift evolution of the Hubble parameter $H(z)$. Using Bayesian inference and MCMC techniques, we constrained the model parameters ($H_0$, $\Omega_{\mathrm{m}0}$, $\alpha$, $\beta$, $\gamma$, and $M$) with high precision, leveraging late-time observational datasets including the CC, PPS, BAO$_1$ and DESI BAO (DR1 and DR2) measurements. The contour plots in Figure \ref{FIG: contour} illustrate the 1$\sigma$ and 2$\sigma$ uncertainty regions for these parameters, demonstrating the model's consistency across datasets, with $H_0$ values ranging from $69.144_{-0.547}^{+0.450}$ to $71.411_{-1.234}^{+1.214}$\kms, closely aligning with recent observational bounds. Figure \ref{FIG: q} reveals a smooth transition from a decelerating to an accelerating phase, with transition redshifts $z_t$ between 0.834 and 0.846, consistent with observational constraints. The current value of the deceleration parameter $q_0$ is estimated to be between $-0.494$ and $-0.549$. Similarly, the EoS parameter is observed to range from $-0.664$ to $-0.693$. These findings are consistent with results from Planck 2018 and WMAP combined with CMB analyses, as illustrated in Figure \ref{FIG: eos}.

    For the CC+PPS dataset, the $f(T, T_\mathcal{G})$ model achieves a best-fit $\chi^2_{\text{min}} = 1661.412$, with $\Delta \text{AIC} = -17.070$ and $\Delta \text{BIC} = -6.155$. The notably negative values of these criteria indicate a strong statistical preference for the $f(T, T_{\mathcal{G}})$ scenario compared to the standard $\Lambda$CDM model. This is demonstrated in Table~\ref{Table: statistical values}, which shows consistently negative $\Delta \mathrm{AIC}$ and $\Delta \mathrm{BIC}$ values, suggesting a statistically better fit across most cosmological datasets. However, the near-zero $\Delta \mathrm{BIC}$ for BAO$_1$ and DESI DR1 highlights the need to refine the model's additional parameters to better balance complexity and fit. Figure~\ref{FIG: whisker_plot} reveals a persistent discrepancy in Hubble constant measurements, with the $f(T, T_{\mathcal{G}})$ gravity model produces $H_0$ estimates ($69-71.5$\kms) that are intermediate between early-Universe constraints (e.g., Planck 2018: $67.4 \pm 0.5$\kms) and late-time observations (e.g., SH0ES: $73.4 \pm 2.0$\kms), partially alleviating but not fully resolving the Hubble tension. However, the $3.2\sigma$-$4.1\sigma$ tension between Planck and SH0ES persists, indicating that while geometric modifications in $f(T, T_\mathcal{G})$ gravity reduce the Hubble discrepancy, they do not fully resolve it, suggesting the need for additional physics or refined observational calibrations. Additionally, the rapid decay of both Hubble and matter perturbations indicates a stable cosmological evolution, potentially suppressing late-time structure formation compared to $\Lambda$CDM. This contrast implies that the model may suppress the growth of large-scale structure, potentially leading to lower matter clustering than observed. Such suppression could be in tension with data from galaxy surveys and weak lensing, which require sufficient structure formation. Therefore, while rapid decay can help alleviate issues like excess power on small scales, it must still be consistent with large-scale structure observations. While the $f(T, T_{\mathcal{G}})$ model effectively describes late-time cosmic acceleration and aligns with low-redshift observations, its deviations from $\Lambda$CDM at higher redshifts suggest it may not fully capture the matter-dominated era. Overall, the model shifts the inferred value of $H_0$ toward higher values relative to the corresponding late-time $\Lambda$CDM baseline, which may reduce---but does not fully eliminate---the discrepancy between early- and late-Universe determinations. For that reason, we interpret the result as a partial alleviation rather than a complete resolution of the Hubble tension.
    
    A full assessment against early-Universe observables, such as CMB anisotropies and BBN constraints, would require extending the model to a complete treatment of recombination-era perturbations, sound-horizon physics, and primordial expansion. At the background level, the model of Eq.~\eqref{Eq: main ode} admits the same effective scaling order for the $T^2$ and $T_\mathcal{G}$ contributions in FLRW cosmology, which suggests that standard radiation- and matter-dominated regimes need not be spoiled. Even so, a full early-Universe consistency analysis remains an open task and should be addressed separately. Future work can leverage upcoming observational probes to rigorously test the $f(T, T_{\mathcal{G}})$ model. The Rubin Observatory's Legacy Survey of Space and Time (LSST) will provide high-precision weak lensing and galaxy clustering data, enabling constraints on the growth rate and clustering amplitude $S_8$, which are sensitive to the suppressed perturbation growth of the model.

\section*{Acknowledgments}
The authors acknowledge that the Ministry of Higher Education, Research, and Innovation (MoHERI) supported this research work through the project BFP/GRG/CBS/24/035. The authors are also thankful to the UoN administration for the continuous support and encouragement for the research work. 
\section*{Data Availability}
There are no associated data with this article. No new data were generated or analysed in support of this research.  

\section*{Analytical Expressions for Perturbation Coefficients}

 \label{Appendix: B}
    In this Appendix, we present the explicit analytical expressions for the coefficients appearing in equation~\eqref{eq:perturbation_ode}. The coefficients are given below as follows
    \begin{small}
  \begin{eqnarray}
 &&\hspace{-1.5cm}   c_0 = -6 \bar{H}^2 \Big(2 \Big(\bar{H}^2 \left(-6 \bar{f}_{TT}+96 \bar{f}_{T T_{\mathcal{G}}} \dot{\bar{H}}-4 \bar{f}_{T_{\mathcal{G}}}+648 \bar{f}_{T_{\mathcal{G}} T_{\mathcal{G}}} {\dot{\bar{H}}}^2\right) \nonumber\\  &&\hspace{-1.0cm} + 12 \bar{H}^4 \left(2 \dot{\bar{H}} \left(6 \bar{f}_{TT T_{\mathcal{G}}} + 60 \bar{f}_{T T_{\mathcal{G}} T_{\mathcal{G}}} \dot{\bar{H}} + 65 \bar{f}_{T_{\mathcal{G}} T_{\mathcal{G}}} + 144 \bar{f}_{T_{\mathcal{G}} T_{\mathcal{G}} T_{\mathcal{G}}} {\dot{\bar{H}}}^2\right)-5 \bar{f}_{T T_{\mathcal{G}}}\right) \nonumber\\  &&\hspace{-1.0cm}-96 H_0^6 \left(\bar{f}_{T_{\mathcal{G}} T_{\mathcal{G}}} - 24 \dot{\bar{H}} \left(\bar{f}_{T T_{\mathcal{G}} T_{\mathcal{G}}}+5 \bar{f}_{T_{\mathcal{G}} T_{\mathcal{G}} T_{\mathcal{G}}} \dot{\bar{H}} \right)\right)+288 \bar{H}^5 \ddot{\bar{H}} \Big(\bar{f}_{T T_{\mathcal{G}} T_{\mathcal{G}}} \nonumber\\  &&\hspace{-1.0cm}+ 6 \bar{f}_{T_{\mathcal{G}} T_{\mathcal{G}} T_{\mathcal{G}}} \dot{\bar{H}} \Big)  -3 \bar{f}_{T_{\mathcal{G}}} \dot{\bar{H}} + 360 \bar{H}^3 \bar{f}_{T_{\mathcal{G}} T_{\mathcal{G}}} \ddot{\bar{H}} + 9216 \bar{H}^8 \bar{f}_{T_{\mathcal{G}} T_{\mathcal{G}} T_{\mathcal{G}}} \dot{\bar{H}} + 2304 \bar{H}^7 \nonumber\\  &&\hspace{-1.0cm} \times \bar{f}_{T_{\mathcal{G}} T_{\mathcal{G}}} \ddot{\bar{H}} \Big)-3 \bar{f}_T\Big)\, ,  \\
 &&\hspace{-1.5cm}  c_1 = -1224 \bar{H}^5 \left(\bar{f}_{T T_\mathcal{G}}+17 \bar{f}_{T_\mathcal{G} T_\mathcal{G}} \dot{\bar{H}}\right)+24 \bar{H}^7 \Big(12 \dot{\bar{H}} \left(\bar{f}_{T T_\mathcal{G} T_\mathcal{G}} + 4 \bar{f}_{T_\mathcal{G} T_\mathcal{G} T_\mathcal{G}} \dot{\bar{H}} \right) \nonumber\\  &&\hspace{-1.0cm} + 11 \bar{f}_{T_\mathcal{G} T_\mathcal{G}} \Big)-\bar{H}^3 \bar{f}_{T_\mathcal{G}} +2304 \bar{H}^9 \bar{f}_{T_\mathcal{G} T_\mathcal{G} T_\mathcal{G}} \dot{\bar{H}} + 576 \bar{H}^8 \bar{f}_{T_\mathcal{G} T_\mathcal{G} T_\mathcal{G}} \ddot{\bar{H}}, \\
    &&\hspace{-1.5cm}    c_2 = -864 \bar{H}^6 \bar{f}_{T_\mathcal{G} T_\mathcal{G}} \, , \\
      &&\hspace{-1.5cm}      c_\mathrm{m} = \kappa^2 \bar{\rho}_{\mathrm{m}} .
        \end{eqnarray}
  \end{small}
 
\section{BAO Distance Measurements from Observational Data} \label{Appendix: A}
  \begin{table*} [!htb]
\centering
\begin{tabular}{| c | c | c | c |}
\hline 
 $z_{\rm eff}$ &  Value & Observable  & Reference  \\ 
\hline
$0.81$	& 	$10.75 \pm 0.43$    &  $D_A/r_d$  &~\cite{Abbott_2018_483}\\
\hline
$0.38$	& 	$10.272 \pm 0.135 \pm 0.074$   &  $D_M/r_d$    & \\
$0.51$	& 	$13.378 \pm 0.156 \pm 0.095$ &  $D_M/r_d$   &~\cite{Alam_2017_470}\\
$0.61$	& 	$ 15.449 \pm 0.189 \pm 0.108$    &  $D_M/r_d$   & \\
\hline
$0.698$	& 	$ 17.65 \pm 0.3$    &  $D_M/r_d$   & \cite{Bautista_2020_500}\\
\hline
$1.48$	& 	$  30.21 \pm 0.79$    &  $D_M/r_d$   & \cite{Neveux_2020_499} \\
\hline
$2.3$	& 	$37.77 \pm 2.13$   &  $D_M/r_d$  &		\cite{Bautista_2017_603}\\
\hline
$2.4$	& 	$36.6 \pm 1.2$  &  $D_M/r_d$  &	\cite{Bourboux_2017_608}\\
\hline
$0.15$	& 	$4.473 \pm 0.159$   &  $D_V/r_d$  &	\cite{Ross_2015_449}\\
\hline
$0.44$	& 	$11.548 \pm 0.559$  &  $D_V/r_d$  &	\\
$0.6$	& 	$14.946 \pm 0.680 $  &  $D_V/r_d$  &	\cite{Kazin_2014_441}\\
$0.73$	& 	$16.931 \pm 0.579$   &  $D_V/r_d$  &	\\
\hline
$1.52$	& 	$26.005 \pm 0.995$   &  $D_V/r_d$  &	\cite{Ata_2017_473}\\
\hline
$0.698$	& 	$ 19.77 \pm 0.47$    &  $D_H/r_d$   & \cite{Bautista_2020_500}\\
\hline
$1.48$	& 	$  13.23 \pm 0.47$    &  $D_H/r_d$   & \cite{Neveux_2020_499} \\
\hline
$2.3$	& 	$9.07 \pm 0.31 $  &  $D_H/r_d$  &	\cite{Bautista_2017_603}\\
\hline
$2.4$	& 	$8.94 \pm 0.22$   &  $D_H/r_d$  &	\cite{Abbott_2018_483}\\
\hline
$0.38$	& 	$12044.07 \pm 251.226 \pm 133.002 $  &  $H r_d$ [km/s]  & \\
$0.51$	& 	$ 13374.09 \pm 251.226 \pm 147.78$    &  $H r_d$  [km/s]    & ~\cite{Alam_2017_470}\\
$0.61$	& 	$ 14378.994 \pm 266.004 \pm 162.558 $    &  $H r_d$ [km/s]   & \\
\hline
\end{tabular}
\caption{Constraints on distance bounds derived from BAO measurements across various observational probes and surveys. The table presents the effective redshift of each measurement, the mean value with its standard deviation, the corresponding observable, and the associated reference.}
\label{Table:BAO1}
\end{table*}

\begin{table*}[!htb]
    \centering
    \begin{tabular}{| c | c | c | c | c |}
        \hline
        Tracer & $z_{\mathrm{eff}}$ & $D_M/r_d$ & $D_H/r_d$ & $r \text{ or } D_V/r_d$ \\
        \hline
        BGS & 0.295 & - & - & $7.93 \pm 0.15$ \\
        LRG1 & 0.510 & $13.62 \pm 0.25$ & $20.98 \pm 0.61$ & $-0.445$ \\
        LRG2 & 0.706 & $16.85 \pm 0.32$ & $20.08 \pm 0.60$ & $-0.420$ \\
        LRG3+ELG1 & 0.930 & $21.71 \pm 0.28$ & $17.88 \pm 0.35$ & $-0.389$ \\
        ELG2 & 1.317 & $27.79 \pm 0.69$ & $13.82 \pm 0.42$ & $-0.444$ \\
        QSO & 1.491 & - & - & $26.07 \pm 0.67$ \\
        Lya QSO & 2.330 & $39.71 \pm 0.94$ & $8.52 \pm 0.17$ & $-0.477$ \\
        \hline
    \end{tabular}
    \caption{Statistical data pertaining to the DESI samples used for the measurements associated with the DESI DR1 BAO. The table presents the effective redshift ($z_{\mathrm{eff}}$) for each observation, along with the corresponding distance ratios: either the correlated pair $(D_M/r_d,\,\, D_H/r_d)$ with correlation coefficient $r$, or the spherically averaged distance $D_V/r_d$.}
    \label{Table:BAO stat DR1}
\end{table*}

\begin{table*}[!htb]
\centering
    \begin{tabular}{| c | c | c | c | c |}
        \hline
        Tracer & $z_{\mathrm{eff}}$ & $D_M/r_d$ & $D_H/r_d$ & $r \text{ or } D_V/r_d$ \\
        \hline
        BGS & 0.295 & - & - & $7.942 \pm 0.075$ \\
        LRG1 & 0.510 & $13.588 \pm 0.167$ & $21.863 \pm 0.425$ & $12.720 \pm 0.099$ \\
        LRG2 & 0.706 & $17.351 \pm 0.177$ & $19.455 \pm 0.330$ & $16.050 \pm 0.110$ \\
        LRG3+ELG1 & 0.934 & $21.576 \pm 0.152$ & $17.641 \pm 0.193$ & $19.721 \pm 0.091$ \\
        ELG2 & 1.321 & $27.601 \pm 0.318$ & $14.176 \pm 0.221$ & $24.252 \pm 0.174$ \\
        QSO & 1.484 & $30.512 \pm 0.760$ & $12.817 \pm 0.516$ & $26.055 \pm 0.398$ \\
        Lya & 2.330 & $38.988 \pm 0.531$ & $8.632 \pm 0.101$ & $31.267 \pm 0.256$ \\
        LRG3 & 0.922 & $21.648 \pm 0.178$ & $17.577 \pm 0.213$ & $19.656 \pm 0.105$ \\
        ELG1 & 0.955 & $21.707 \pm 0.335$ & $17.803 \pm 0.297$ & $20.008 \pm 0.183$ \\
        \hline
    \end{tabular}
    \caption{Statistical data pertaining to the DESI samples used for the measurements associated with the DESI DR2 BAO. The table presents the effective redshift ($z_{\mathrm{eff}}$) for each observation, along with the corresponding distance ratios: either the correlated pair $(D_M/r_d,\,\, D_H/r_d)$ with correlation coefficient $r$, or the spherically averaged distance $D_V/r_d$.}
    \label{Table:BAO stat DR2}
\end{table*}


\providecommand{\href}[2]{#2}\begingroup\raggedright\endgroup

\end{document}